\documentclass[nobibnotes,superscriptaddress,showpacs,preprintnumbers,twocolumn]{revtex4}

\usepackage{amsmath,amssymb,graphicx}

\begin{document}

\title{Resonant leptogenesis and tribimaximal leptonic mixing with $A_4$ symmetry}

\author{G. C. Branco}
\email{gbranco@ist.utl.pt}
\affiliation{Departamento de F\'{\i}sica and Centro de F\'{\i}sica Te\'orica de Part\'{\i}culas (CFTP),
Instituto Superior T\'ecnico, Avenida Rovisco Pais, 1049-001 Lisboa, Portugal}

\author{R.~Gonz\'{a}lez Felipe}
\email{gonzalez@cftp.ist.utl.pt}
\affiliation{Area Cient\'{\i}fica de F\'{\i}sica, Instituto Superior de Engenharia
de Lisboa, Rua Conselheiro Em\'{\i}dio Navarro 1, 1959-007 Lisboa, Portugal}
\affiliation{Departamento de F\'{\i}sica and Centro de F\'{\i}sica Te\'orica de
Part\'{\i}culas (CFTP), Instituto Superior T\'ecnico, Avenida Rovisco Pais, 1049-001 Lisboa, Portugal}

\author{M. N. Rebelo}
\email{rebelo@ist.utl.pt}
\affiliation{Departamento de F\'{\i}sica and Centro de F\'{\i}sica Te\'orica de Part\'{\i}culas (CFTP),
Instituto Superior T\'ecnico, Avenida Rovisco Pais, 1049-001 Lisboa, Portugal}

\author{H.~Ser\^odio}
\email{hserodio@cftp.ist.utl.pt}
\affiliation{Departamento de F\'{\i}sica and Centro de F\'{\i}sica Te\'orica de Part\'{\i}culas (CFTP),
Instituto Superior T\'ecnico, Avenida Rovisco Pais, 1049-001 Lisboa, Portugal}

\begin{abstract}
We investigate the viability of thermal leptogenesis in type-I seesaw models with leptonic flavour symmetries that lead to tribimaximal neutrino mixing. We consider an effective theory with an $A_4 \times Z_3 \times Z_4$ symmetry, which is spontaneously broken at a scale much higher than the electroweak scale. At the high scale, leptonic Yukawa interactions lead to exact tribimaximal mixing and the heavy Majorana neutrino mass spectrum is exactly degenerate. In this framework, leptogenesis becomes viable once this degeneracy is lifted either by renormalization group effects or by a soft breaking of the $A_4$ symmetry. The implications for low-energy neutrino physics are discussed.
\end{abstract}
\pacs{ }
\maketitle

\section{Introduction}
\label{sec1}
Fermion masses and mixing have become even more puzzling with the recent discovery of neutrino masses and large leptonic mixing. One of the approaches often adopted in the search for a possible solution for the flavour puzzle consists of the introduction of family symmetries which constrain the flavour structure of Yukawa couplings and lead to predictions for fermion masses and  mixings. Harrison, Perkins and Scott (HPS) \cite{Harrison:2002er} have pointed out that leptonic mixing at low energies could be  described by the so-called tribimaximal mixing matrix, which is a good representation of the present data within $1\sigma$. The special form of this matrix is suggestive of a symmetry related to possible subgroups of $SU(3)$ \cite{Luhn:2007yr}. This fact prompted many attempts at finding an underlying symmetry leading to this special pattern of mixing~\cite{Altarelli:2007gb}. Of particular interest are models based on $A_4$ symmetry, which was first introduced~\cite{Wyler:1979fe} as a possible family symmetry for the quark sector and is now mostly used for the lepton sector~\cite{Ma:2001dn,Altarelli:2005yx,He:2006dk,deMedeirosVarzielas:2005qg,Altarelli:2005yp,Hirsch:2008rp}. In the leptonic sector, neutrino masses are known to be much smaller than the masses of all other fermions and, in addition, leptonic mixing includes large mixing, thus drastically differing from the quark sector. An elegant explanation for the smallness of neutrino masses is the seesaw mechanism \cite{seesaw}, which has also the advantage of providing a simple and attractive leptogenesis mechanism for the generation of the observed baryon asymmetry in the Universe, through the decay of heavy Majorana neutrinos~\cite{Fukugita:1986hr,Davidson:2008bu}. In such a scenario, a relationship between low-energy observables and the size of the leptonic asymmetry can only be established in some special cases
~\cite{Branco:2001pq,Rebelo:2002wj,GonzalezFelipe:2003fi,Branco:2006ce,Davidson:2007va}.

In this paper, we address the question of the viability of leptogenesis in models with leptonic flavour symmetries leading to the HPS mixing matrix in the framework of the seesaw mechanism. Our starting point is an effective Lagrangian with an $A_4 \times Z_3 \times Z_4$ symmetry which is spontaneously broken by the vacuum expectation values (VEV) of $SU(2)_L\times U(1)_Y$ singlet scalar fields at a scale much higher than the electroweak scale. The resulting Yukawa couplings at this high scale correspond to exact HPS mixing with the possibility of Majorana-type $CP$ violation, as well as to exact degeneracy of the heavy Majorana neutrinos. In this model, leptogenesis becomes viable once the exact degeneracy of the heavy Majorana neutrinos is lifted. We analyze two possible different ways of lifting this degeneracy, either radiatively, when renormalization group effects are taken into account, or through a soft breaking of the $A_4$ symmetry. An interesting feature of our model is the fact that the combination of Yukawa couplings appearing in the leptonic $CP$ asymmetries relevant for flavoured leptogenesis \cite{Barbieri:1999ma,Endoh:2003mz,Fujihara:2005pv,Abada:2006fw,Nardi:2006fx,Abada:2006ea}
does not vanish at this high scale. This is a particular feature of our framework.

Our paper is organized as follows. In the next section, we present our framework, indicating the flavour symmetry, together with the matter content of the model. In Sec.~\ref{sec3}, we describe the implications of the flavour symmetry on mixing angles, neutrino mass spectrum and other low-energy observables. Section~\ref{sec4} deals with leptogenesis where we describe two mechanisms to obtain viable leptogenesis in our framework, namely through radiative leptogenesis and through soft breaking of the family symmetry. Our conclusions are summarized in Sec.~\ref{sec5}.

\section{Framework: symmetry and matter content}
\label{sec2}
We work in the framework of an extension of the standard model (SM), consisting of the addition of three right-handed neutrinos. The scalar sector, apart from the usual SM Higgs doublet $\phi$, is extended through the introduction of  four types of heavy scalar fields, $\Phi$, $\Psi$, $\Theta$ and $\chi$, that are singlets under $SU(3)\times SU(2)_L\times U(1)_Y$. Furthermore, we impose an  $A_4 \times Z_3 \times Z_4$ symmetry to the Lagrangian. As is well known, $A_4$ is a discrete symmetry corresponding to the even permutation of four objects having four irreducible representations: three inequivalent one-dimensional representations ($\mathbf{1}$, $\mathbf{1^\prime}$, $\mathbf{1^{\prime\prime}}$) and a three-dimensional representation ($\mathbf{3}$). The following multiplication rules hold: $\mathbf{1^\prime} \otimes \mathbf{1^{\prime \prime}}=\mathbf{1}$, $\mathbf{1^{\prime}} \otimes \mathbf{1^{\prime}}=\mathbf{1^{\prime \prime}}$, $\mathbf{1^{\prime \prime}} \otimes \mathbf{1^{\prime \prime}}=\mathbf{1^{\prime}}$ and $\mathbf{3} \otimes \mathbf{3}=\mathbf{1}\oplus\mathbf{1^\prime}\oplus\mathbf{1^{\prime\prime}}\oplus \mathbf{3_s}\oplus \mathbf{3_a}$. Therefore, the product of two triplets, $a=(a_1,a_2,a_3)$ and $b=(b_1,b_2,b_3)$, yields

\begin{widetext}
\begin{center}
\begin{table}[h]
\caption{\label{reps} Representations of the fields under $A_4 \times Z_3 \times Z_4$ and $SU(2)_L \times U(1)_Y$.}
\begin{ruledtabular}
\begin{tabular}{ccccccccc}
Field &$\ell$&$e_R,\mu_R,\tau_R$&$\nu_R$&$\phi$&$\Phi$&$\Psi$&$\Theta$&$\chi$\\
\hline
$A_4$&$\mathbf{3}$&$\mathbf{1}$, $\mathbf{1^\prime}$,
$\mathbf{1^{\prime\prime}}$&$\mathbf{3}$&$\mathbf{1}$&
$\mathbf{1}$&$\mathbf{3}$&$\mathbf{3}$&$\mathbf{1}$\\
$Z_3$&$\omega$&$\omega$&$\omega$&$1$&$1$&$1$&$1$&$\omega$\\
$Z_4$&$1$&$-1$&$-i$&$1$&$i$&$-1$&$i$&$-1$\\
$SU(2)_L\times U(1)_Y$&$(2,1/2)$&$(1,1)$&$(1,0)$&$(2,-1/2)$&$(1,0)$&$(1,0)$&$(1,0)$&$(1,0)$\\
\end{tabular}
\end{ruledtabular}
\end{table}
\end{center}
\end{widetext}

\begin{align}
\nonumber(a \otimes b)_\mathbf{1}&=a_1b_1+a_2b_2+a_3b_3\,,\\
\nonumber(a \otimes b)_\mathbf{1^\prime}&=a_1b_1+\omega^2 a_2b_2+\omega a_3b_3\,,\\
(a\otimes b)_\mathbf{1^{\prime\prime}}&=a_1b_1+\omega a_2b_2+\omega^2a_3b_3\,,\\
\nonumber(a \otimes b)_\mathbf{3_s}&=
(a_2b_3+a_3b_2,a_3b_1+a_1b_3,a_1b_2+a_2b_1)\,,\\
\nonumber(a\otimes b)_\mathbf{3_a}&=
(a_2b_3-a_3b_2,a_3b_1-a_1b_3,a_1b_2-a_2b_1)\,,
\end{align}
where $\omega$ is the cube root of unity, i.e. $\omega=e^{i 2\pi/3}$. For the symmetric product of three triplets one has
\begin{eqnarray}
(a \otimes b\otimes c)_\mathbf{1}=\sum_{i,j,k}^3a_ib_jc_k\,,&&\text{with $i\neq j\neq k$}\,.
\end{eqnarray}

Table \ref{reps} shows how the various fields transform under the different symmetry groups.
It is clear from Table~\ref{reps} that it is not possible to introduce SM-like Yukawa couplings for the charged leptons since these would break the $A_4$ as well as the $Z_4$ symmetries. Similar couplings for the neutral leptons are forbidden by the $Z_4$ symmetry. Majorana mass terms for the right-handed neutrinos are not allowed, but a Yukawa-type interaction term can be built with the $A_4$ singlet field $\chi$.
Direct couplings of the right-handed neutrinos to $\Phi$, $\Psi$ and $\Theta$ are also forbidden by the discrete symmetries. It is necessary to introduce higher dimensional operators to get nonzero charged-lepton masses and to allow for the generation of Dirac mass terms for the neutrinos. We assume that above a cutoff scale $\Lambda$ there is unknown physics, which for scales below $\Lambda$ is expressed in terms of higher dimensional operators. The scale at which $A_4 \times Z_4$ is broken is assumed to be lower than the cutoff $\Lambda$, but still close to it. On the other hand, the breaking of $Z_3$, being responsible for the heavy Majorana neutrino masses, can occur at a much lower scale.

This gives rise to the following effective ($d\leq 5$) Lagrangian terms :
\begin{equation}\label{LagraY}
\begin{split}
&\frac{y_1^\ell}{\Lambda}\left(\bar{\ell}\,\Psi\right)_{\mathbf{1}}\phi\,e_R+
\frac{y_2^\ell}{\Lambda}\left(\bar{\ell}\,\Psi\right)_{\mathbf{1^{\prime\prime}}}\phi\,\mu_R + \frac{y_3^\ell}{\Lambda}\left(\bar{\ell}\,\Psi\right)_{\mathbf{1^\prime}}\phi\, \tau_R + \\ &\frac{y_1^\nu}{\Lambda}\left(\bar{\ell}\,\nu_R\right)_{\mathbf{1}}\Phi\,\tilde{\phi} + \frac{y_2^\nu}{\Lambda}\left(\bar{\ell}\,\nu_R\,\Theta\right)_{\mathbf{1}}\tilde{\phi}+\frac12 y_R \chi\left(\overline{\nu_R^c}\nu_R\right)_{\mathbf{1}} + \text{H.c.}\,
\end{split}
\end{equation}

We do not impose $CP$ invariance, so in this model $CP$ is violated at the Lagrangian level. We assume that there is a region of the parameter space of the scalar potential where the heavy scalars develop VEV of the form
\begin{equation} \label{vevalign}
\begin{array}{ccc}
\left<\Phi\right>=u\,,\quad&\left<\Psi\right>=\left(s,s,s\right)\,,
\quad&\left<\Theta\right>=\left(0,t,0\right)\,,
\end{array}
\end{equation}
thus breaking the $A_4\times Z_4$ symmetry. The $Z_3$ symmetry is only broken when the singlet field $\chi$ develops a VEV. Needless to say that the choice of VEV directions in Eq.~(\ref{vevalign}) requires a stable vacuum alignment of the triplet fields $\Psi$ and $\Theta$. Yet, the presence of terms like $\left(\Psi^\dagger\Psi\right)_\mathbf{3_s}\left(\Theta^\dagger\Theta\right)_\mathbf{3_s}$, $\left(\Psi^\dagger\Theta^\dagger\right)_\mathbf{1^\prime}\left(\Psi\Theta\right)_\mathbf{1^{\prime\prime}}$ and $\left(\Psi^\dagger\Theta^\dagger\right)_\mathbf{3_a}\left(\Psi\Theta\right)_\mathbf{3_a}$ would clearly distinguish between the different vacuum directions. Such an alignment can be naturally achieved for instance in supersymmetric dynamical completions~\cite{Altarelli:2005yx,He:2006dk,deMedeirosVarzielas:2005qg} or in the presence of extra dimensions~\cite{Altarelli:2005yp}.

The effective Lagrangian will then lead to the following Yukawa-type couplings and direct mass terms:
\begin{widetext}
\begin{equation}
\begin{split}\label{LagraY2}
-\mathcal{L}_Y^{eff}=&f_1^s\,\left(\overline{\ell_e}+\overline{\ell_\mu}+
\overline{\ell_\tau}\right)\phi\, e_R+f_2^s\,\left(\overline{\ell_e}+\omega\,
\overline{\ell_\mu}+\omega^2\,\overline{\ell_\tau}\right)\phi\, \mu_R
 + f_3^s\, \left(\overline{\ell_e}+\omega^2\,\overline{\ell_\mu}+\omega\,
 \overline{\ell_\tau}\right)\phi\, \tau_R \\
 &+ f^u\,\left(\overline{\ell_e}\,\nu_{1R}+
\overline{\ell_\mu}\,\nu_{2R} +  \overline{\ell_\tau}\,\nu_{3R}\right)\tilde{\phi}+ f^t\,\left(\overline{\ell_e}\,\nu_{3R}+
\overline{\ell_\tau}\,\nu_{1R}\right)\tilde{\phi} +
M\left(\overline{\nu_{1R}^c}\nu_{1R}+ \overline{\nu_{2R}^c}\nu_{2R}+
\overline{\nu_{3R}^c}\nu_{3R}\right)+\text{H.c.} \\
=&Y^\ell_{ij}\,\overline{\ell_i}\,\phi\,e_{j\,R}+Y^\nu_{ij}\,
\overline{\ell_i}\,\tilde{\phi}\,\nu_{j\,R}+M_R^{ij}\,\overline{\nu_{i\,R}^{c}}\,
\nu_{j\,R}+\text{H.c.}\,,
\end{split}
\end{equation}
\end{widetext}
where $M=y_R \langle \chi \rangle$ and the following definitions have been introduced:
\begin{align}
f_i^s\equiv\frac{s}{\Lambda}\,y_i^\ell\,,\quad f^u\equiv\frac{u}{\Lambda}
\,y_1^\nu,\quad f^t\equiv\frac{t}{\Lambda}\,y_2^\nu \quad (i=1,2,3)\,.
\end{align}
These effective Yukawa couplings are assumed to be within the perturbative regime, i.e. $f^{\left(s,u,t\right)} \lesssim 1$. The effective Yukawa couplings  and Majorana mass matrix are of the form
\begin{align}\label{matrix}
Y^\ell=
\begin{pmatrix}
f_1^s&f_2^s&f_3^s\\
f_1^s&\omega\,f_2^s&\omega^2\,f_3^s\\
f_1^s&\omega^2\,f_2^s&\omega\,f_3^s
\end{pmatrix},
\end{align}
\begin{align}\label{ynu}
Y^\nu=e^{i\alpha_1}\dfrac{\sqrt{M}}{v}
\begin{pmatrix}
x& 0 &y\,e^{i\alpha}\\
0 &x& 0\\
y\,e^{i\alpha}& 0 &x
\end{pmatrix},
\end{align}
\begin{align}
M_R =\begin{pmatrix}
M&&\\
&M&\\
&&M
\end{pmatrix},
\end{align}
where $v$ denotes the vacuum expectation value of the usual SM Higgs doublet, $\langle\phi^0\rangle=v$,
and $x$ and $y$ stand for the real and positive quantities
\begin{align}\label{xy}
x\equiv\frac{v }{\sqrt{M}}\left| f^u\right|\quad&\text{and}\quad y\equiv
\frac{v }{\sqrt{M}}\left| f^t\right|\,.
\end{align}

The phases $\alpha_1$ and $\alpha_2$ are the arguments of $f^u$ and $f^t$, respectively, and $\alpha\equiv \alpha_2-\alpha_1$ is the only physical phase remaining in  $Y^\nu$, since the global phase $\alpha_1$, factored out in Eq.~(\ref{ynu}), can be rotated away. Similarly, the phases in $f_i^s$ can be eliminated through the rephasing of the $e_{i\,R}$ fields. Therefore, there is no loss of generality in working with real ${f_i}^s$ and with the only phases remaining in  $Y^\ell$ due to $\omega$ and $\omega^2$. We shall see that the phase $\alpha$ together with the phase in $\omega$ are the only phases which violate $CP$.

The neutrino Yukawa matrix can be rewritten as
\begin{align} \label{Ynu}
Y^\nu=V\,K^{1/2}\,\left|d_D\right|\,K^{1/2}\,V^T\,,
\end{align}
with
\begin{align}\label{matrixV}
V=\begin{pmatrix}
\frac{1}{\sqrt{2}}& 0 &-\frac{1}{\sqrt{2}}\\
0 &1& 0\\
\frac{1}{\sqrt{2}}& 0 &\frac{1}{\sqrt{2}}
\end{pmatrix}\,,
\end{align}
\begin{align}
d_D=\frac{\sqrt{M}}{v}\,\text{diag}\left(x+y\,e^{i\alpha},x,x-y\,
e^{i\alpha}\right)=\left|d_D\right|\,K\,,
\end{align}
and
\begin{align}\label{majph}
K&=\text{diag}\left(e^{i\sigma_1},1\,,e^{i\sigma_2}\right)\,,\nonumber\\
\sigma_1&=\text{arg}\left(x+y\,e^{i\alpha}\right)\,,\; \sigma_2=\text{arg}\left(x-y\,e^{i\alpha}\right)\,.
\end{align}
For the charged-lepton Yukawa matrix we can write $Y^\ell=U_\omega\,d_\ell$ with $d_\ell=\sqrt{3}\,\text{diag}\,(f_1^s,f_2^s,f_3^s)$ and
\begin{align}
U_\omega=\frac{1}{\sqrt{3}}
\begin{pmatrix}
1&1&1\\
1&\omega&\omega^2\\
1&\omega^2&\omega
\end{pmatrix}.
\end{align}

\section{Low-energy observables}
\label{sec3}
Since we work in the seesaw framework, the heavy Majorana neutrino mass scale $M$ is assumed to be much higher than the electroweak scale and the masses of the light neutrinos are simply given by the well-known effective mass matrix
\begin{equation}\label{mnu}
m_\nu =-m_D\,M_R^{-1}\,m_D^T\,,
\end{equation}
where $m_D$ is the Dirac-type neutrino mass matrix in the weak basis (WB) where the charged-lepton mass matrix is diagonal and real,
\begin{equation}
m_D=v\,U_\omega^\dagger Y^\nu.
\label{mdu}
\end{equation}
From Eqs.(\ref{Ynu}), (\ref{mnu}) and (\ref{mdu}) we then find
\begin{equation}
\begin{split}
m_\nu &=-v^2\,U_\omega^\dagger \, Y^\nu\,M_R^{-1}\,Y^{\nu T} U_\omega^\ast \\
&=U_\omega^\dagger V\,K^\prime\,
\left|d_\nu\right|\,K^\prime\,V^T\, U_\omega^\ast \,,
\end{split}
\end{equation}
where $K^\prime=e^{i\pi/2}\,K$ and
\begin{align}
\left|d_\nu\right|=\frac{v^2}{M}
\left|d_D\right|^2\equiv\text{diag}\left(m_1,m_2,m_3\right).
\end{align}
The Pontecorvo-Maki-Nakagawa-Sakata (PMNS) leptonic mixing matrix at low energies is thus given by
\begin{widetext}
\begin{align}\label{PMNS}
U_\text{PMNS}=U_\omega^\dagger V\,K^\prime=e^{i\,\left(\sigma_1+\pi/2\right)}\begin{pmatrix}
1&&\\
&\omega^2&\\
&&\omega
\end{pmatrix}
\begin{pmatrix}
\sqrt{\frac{2}{3}}&\frac{1}{\sqrt{3}}&0\\
-\frac{1}{\sqrt{6}}&\frac{1}{\sqrt{3}}&-\frac{1}{\sqrt{2}}\\
-\frac{1}{\sqrt{6}}&\frac{1}{\sqrt{3}}&\frac{1}{\sqrt{2}}
\end{pmatrix}
\begin{pmatrix}
1&&\\
&e^{i\beta_1}&\\
&&e^{i\beta_2}
\end{pmatrix},
\end{align}
\end{widetext}
with $\beta_1 = -\sigma_1$ and $\beta_2 = \sigma_2-\sigma_1$. We remark that the phases factored out to the left have no physical meaning, since they can be eliminated by a redefinition of the physical charged-lepton fields. Therefore, the only phases appearing in $U_\text{PMNS}$ are the Majorana phases $\beta_1$ and $\beta_2$. After factoring out the additional Majorana-type $CP$ violating phases, this mixing matrix coincides with the HPS matrix. The zero entry in $U_\text{PMNS}$ implies that there is no Dirac-type $CP$ violation.

In the limit of vanishing $\alpha$, there is no $CP$ violation in $U_\text{PMNS}$. However, the remaining phase of $\omega$, entering in $m_D$, does imply $CP$ violation at high energies. This can be seen by recalling~\cite{Branco:2001pq} that, in this class of models, the necessary and sufficient condition for having $CP$ invariance is that in the WB where $m_\ell$ and $M_R$ are diagonal and real, the condition $\arg (m_D)_{ij} = \beta_i/2 - (2p_j+1)\pi/4$ is satisfied with arbitrary $\beta_i$ and integer numbers $p_j\,$. It can be readily verified that the matrix $m_D$ given by Eq.~(\ref{mdu}) does not satisfy this condition even for $\alpha = 0$.

The light neutrino masses $m_1$, $m_2$ and $m_3$ are given by
\begin{align}\label{masses}
m_1& =x^{ 2}+y^{2}+2x y\cos\alpha\,,\nonumber\\
m_2& = x^2\,,\\
m_3& =x^{2}+y^{2}-2x y\cos\alpha\,.\nonumber
\end{align}
The three charged-lepton masses are determined by the three Yukawa couplings $f_i^s$. Note that the present model is highly constrained. The nine physical quantities consisting of the three light neutrino masses, the three mixing angles and three $CP$-violating phases (contained in a general $U_\text{PMNS}$ matrix) are entirely fixed in terms of three real parameters, namely, $x$, $y$ and $\alpha$.

One has the following constraints on the mixing matrix $U_\text{PMNS}$ and the light neutrino masses:
\begin{itemize}
  \item[(i)] The mixing angles are entirely fixed by the  $A_4 \times Z_4 \times Z_3$ symmetry,
leading to the HPS structure at the scale of the breaking of this symmetry and, consequently, predicting no Dirac-type $CP$ violation;
 \item[(ii)] The remaining five physical quantities $\beta_1$, $\beta_2$, $m_1$,
$m_2$ and $m_3$, are determined by the three parameters $x$, $r \equiv x/y$,
and $\alpha$ through Eqs.~(\ref{majph}) and (\ref{masses}).
\end{itemize}

We shall see that only a normal neutrino ordering is allowed in this model and, furthermore, the two existing experimental constraints, to wit the two neutrino mass-squared differences, strongly correlate the allowed values for the parameters $r$ and $\cos \alpha$. The knowledge of the absolute neutrino mass scale would fix $x$.

Clearly, the relations written in this subsection would be exact provided that there was no running of the coefficients defined at the scale of $A_4\times Z_4$ symmetry breaking. Yet the light neutrino masses and the charged-lepton masses are only generated after spontaneous symmetry breakdown, when the field $\phi$ acquires a VEV. In particular, the zero entry in $U_\text{PMNS}$ is not exact. Such deviations are, however, negligibly small.

In order to see how the experimental knowledge on the neutrino mass spectrum constrains the allowed parameter space, let us recall the following experimental constraints  at $2\,\sigma$ confidence level~\cite{Schwetz:2008er}:
\begin{align}\label{experimental}
\Delta m^2_\text{atm}&\equiv\left|m_3^2-m_2^2\right|=\left(2.18 - 2.64\right)\times 10^{-3}\,
\text{eV}^2\,,\nonumber\\
\Delta m^2_\text{sol}&\equiv \;m_2^2-m_1^2\;=
\left(7.25 - 8.11\right)\times 10^{-5}\,\text{eV}^2\,,
\end{align}
with the best-fit values~\cite{Schwetz:2008er}
\begin{align}\label{expbestfit}
\left(\Delta m^2_\text{atm}\right)_{\text{best fit}}&=2.40\times10^{-3}\,\text{eV}^2\,,\nonumber\\
\left(\Delta m^2_\text{sol}\right)_{\text{best fit}}&=7.65\times10^{-5}\,\text{eV}^2.
\end{align}
The sign of $(m_3 - m_2)$ is dictated by the ordering of the neutrino masses, i.e. positive for normal ordering and negative for inverted ordering.

\begin{figure*}[t]
\begin{tabular}{cc}
\includegraphics[width=7.2cm]{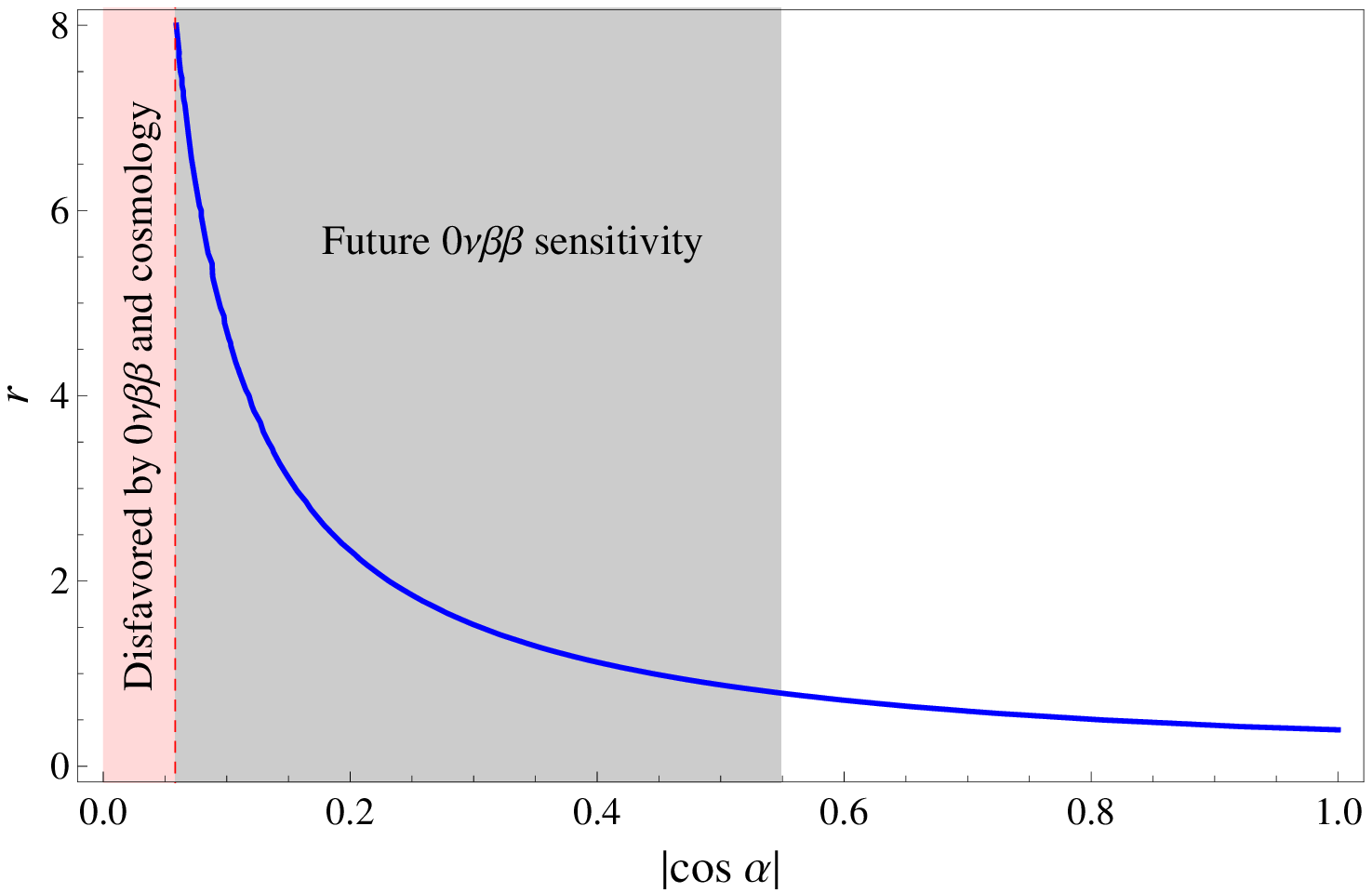}&
\includegraphics[width=7.5cm]{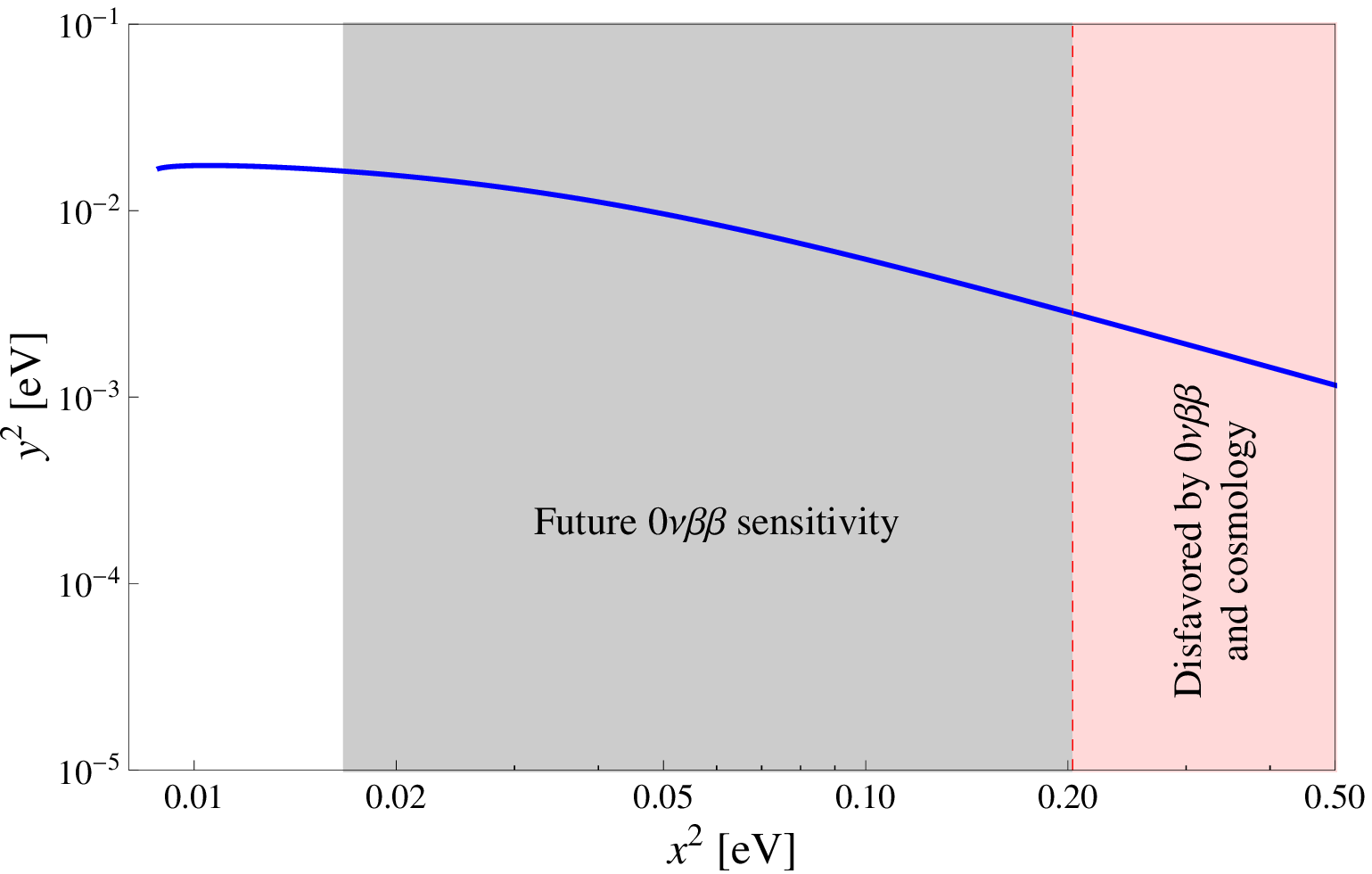}
\end{tabular}
\caption{\label{figure1} (color online). The parameter region allowed by the model: the ratio $r \equiv y/x$
as a function of $\cos\,\alpha$ (left plot) and $y^2$ as a function of $x^2 \equiv m_2$ (right plot).}
\end{figure*}

Let us first consider the case of normal ordering, with $m_3 > m_2$. In this case, one obtains from Eq.~(\ref{masses}) the following constraint:
\begin{align}\label{aaa}
y -2\,x \cos\alpha\, >0,
\end{align}
while the condition  $m_2 > m_1$ leads to
\begin{align}\label{bbb}
y + 2\,x \cos\alpha\, <0.
\end{align}
It is clear that Eqs.~(\ref{aaa}) and (\ref{bbb}) can only be satisfied if $\cos\alpha\, <0$, since $x$ and $y$ are positive. Thus, normal ordering requires the parameter $\alpha$ to be in the second or third quadrant.

Similar considerations applied to the case of inverted ordering, implying
\begin{align}\label{ccc}
y -2\,x \cos\alpha\, <0,
\end{align}
together with Eq.~(\ref{bbb}), since one would still require $m_2 > m_1$. Since Eqs.~(\ref{bbb}) and (\ref{ccc}) cannot be simultaneously verified, one concludes that the present model does not accommodate an inverted ordering for the neutrino mass spectrum.

\begin{figure*}[t]
\begin{tabular}{cc}
\includegraphics[width=7.5cm]{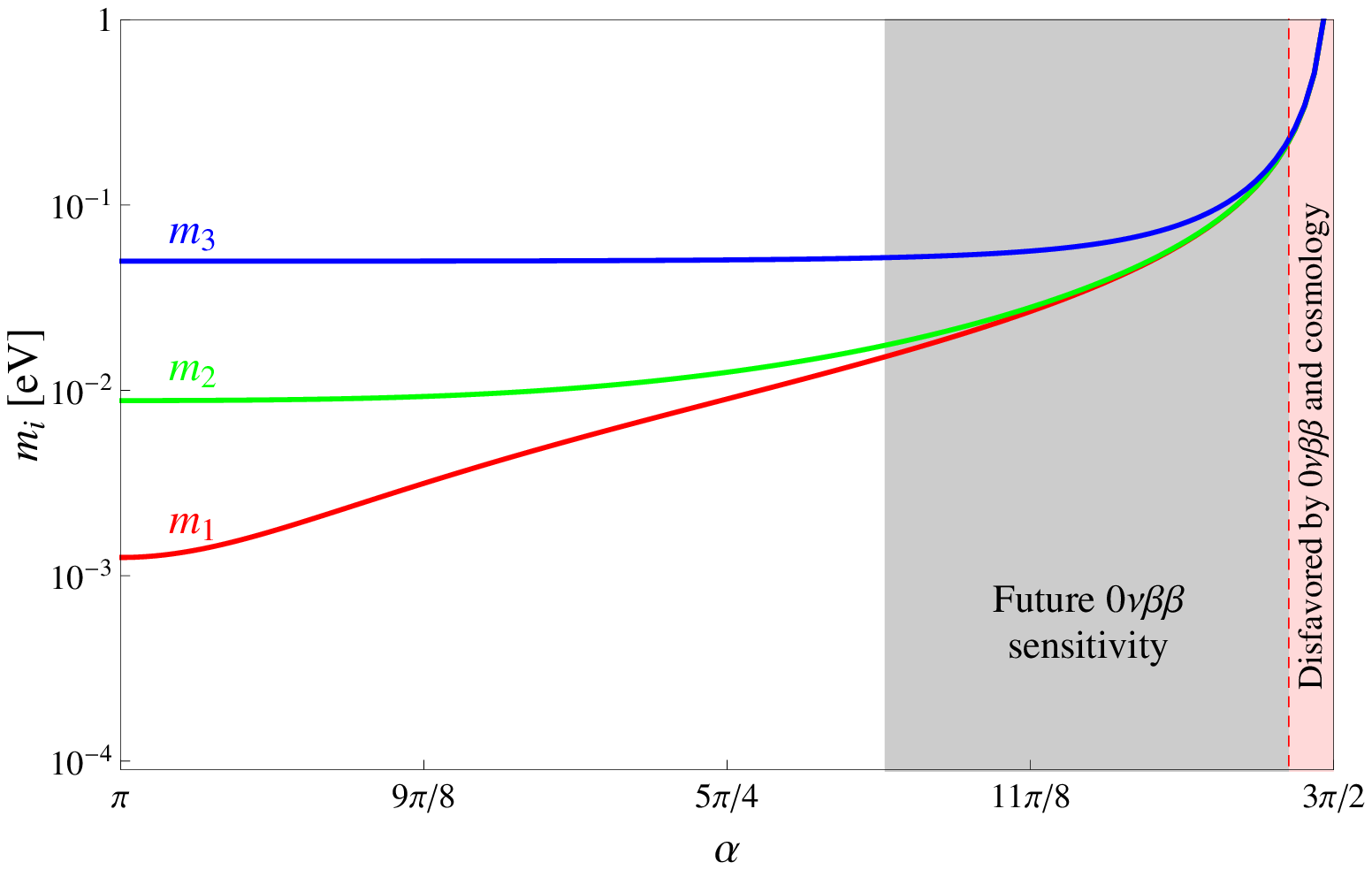}&
\includegraphics[width=7.5cm]{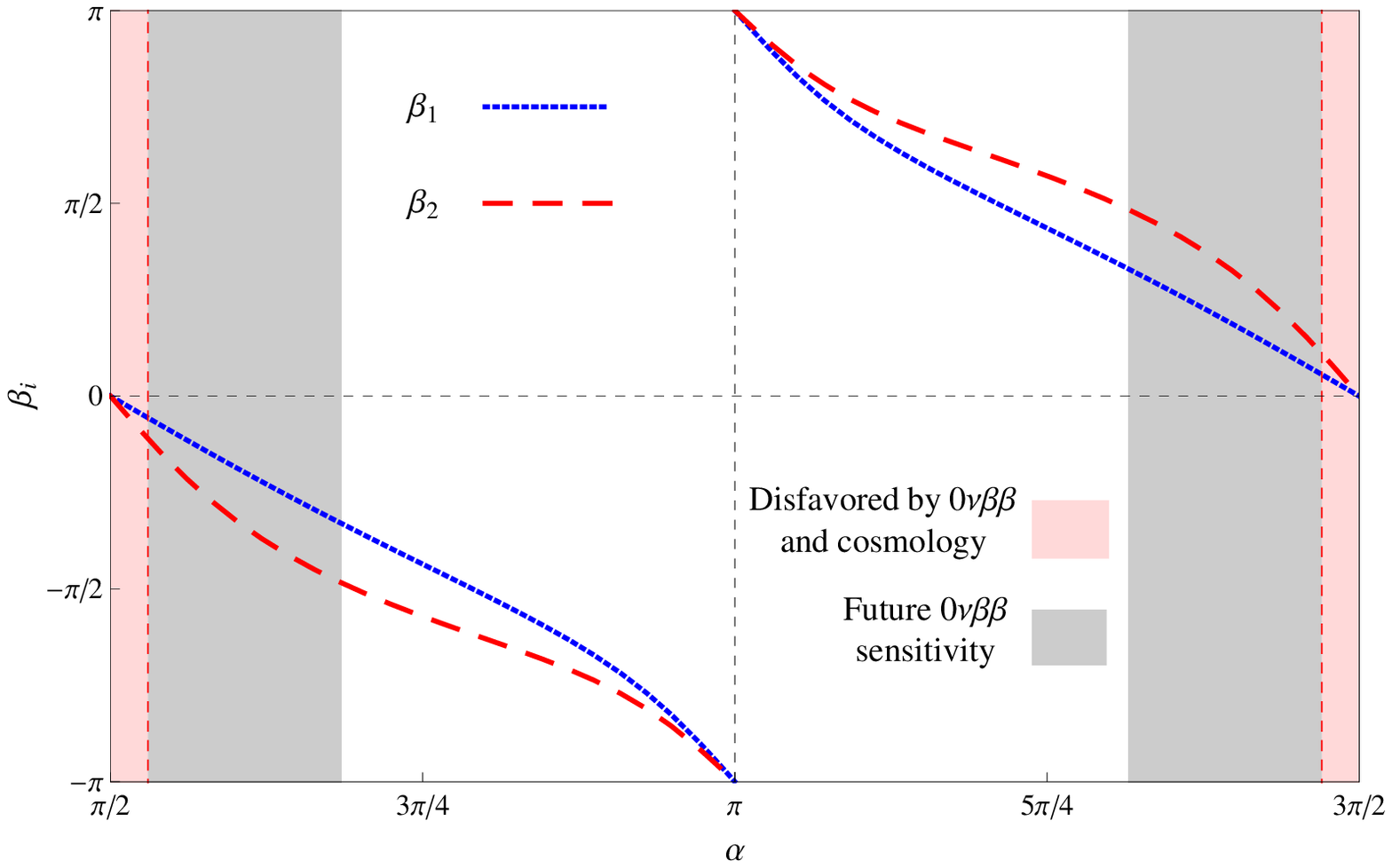}
\end{tabular}
\caption{\label{figure2} (color online).  Low-energy neutrino parameters. On the left plot, the light neutrino
masses $m_i$ are plotted as a function of the high-energy phase $\alpha$, while on the right plot,
the low-energy $CP$-violating Majorana phases $\beta_i$ are displayed as a function of the phase $\alpha$.}
\end{figure*}

The ratio $\Delta m^2_\text{sol}/\Delta m^2_\text{atm}$ also implies a strong correlation between the allowed values for $r$ and $\cos \alpha$.
Indeed, from Eqs.~(\ref{masses}) and (\ref{experimental}) one obtains
\begin{equation}
\frac{\Delta m^2_\text{sol}}{\Delta m^2_\text{atm}} =
\frac{1 + 2r(r-|\cos \alpha|)}{1 + 2r(r+|\cos \alpha|)} \,
\frac{(1 - 2r|\cos \alpha|)}{(1 + 2r|\cos \alpha|)}\,,
\label{ratio}
\end{equation}
where we have taken into account that $\cos \alpha < 0 $. This correlation is presented in Fig.~\ref{figure1} (left plot), for the best-fit values of the solar and atmospheric data given in Eq.~(\ref{expbestfit}). Hereafter, we only use these central values since their experimental dispersion would only contribute to a small enlargement of the allowed region. The light (red) shaded area is currently disfavoured by cosmological observational data. The recent WMAP five-year data~\cite{Komatsu:2008hk} alone constrains the sum of light neutrino masses below 1.3 eV. When combined with baryonic acoustic oscillation and type-Ia supernova data this bound is more restrictive, $\sum_i\,m_i<0.61$~eV. In Fig.~\ref{figure1} we also show the $(x,y)$ parameter region allowed by the model (right plot). This region has a lower bound for $x^2$ and an upper bound for $y^2$ that can be easily understood through the use of the relation
\begin{align}\label{atmsol}
m_3^2+m_1^2-2m_2^2=\Delta m^2_\text{atm}-\Delta m^2_\text{sol}\,.
\end{align}
Clearly, $x^2=\left(m_1^2+\Delta m^2_\text{sol}\right)^{1/2} \gtrsim (\Delta m^2_\text{sol})^{1/2} \simeq 8.7\times 10^{-3}$~eV. This lower limit corresponds to $\alpha\sim\pi$. Moreover, in this limit $m_1^2 \sim (x-y)^4$ is very small when compared with $m_3^2 \sim (x+y)^4$ and, therefore, one has
 \begin{align}\label{y2}
 y \simeq \left|(\Delta m^2_\text{atm})^{1/4}-x\right|
 \lesssim (\Delta m^2_\text{atm})^{1/4}-(\Delta m^2_\text{sol})^{1/4} ,
 \end{align}
implying $ y^2 \lesssim 1.6\times 10^{-2}$~eV. The corresponding light neutrino masses are plotted in Fig.~\ref{figure2} as a function of the phase $\alpha$. Since their dependence on $\alpha$ is expressed only in terms of $\cos\alpha$, we only need to analyze one quadrant, chosen here to be the third quadrant. The lightest neutrino mass has a lower bound given by
\begin{align}
m_1 \gtrsim\left[2(\Delta m^2_\text{sol})^{1/4}-(\Delta m^2_\text{atm})^{1/4}\right]^2 \simeq 1.2\times10^{-3}\,\text{eV}.
\end{align}
The neutrino mass hierarchy is maximal when $\alpha=\pi$, while an almost degenerate spectrum is obtained for $\alpha \simeq\,\pi/2$ or $\alpha\simeq\, 3\pi/2$. Finally, the cosmological bound restricts the phase $\alpha$ to the range $1.04\, \pi/2 \lesssim \alpha \lesssim 2.96 \,\pi/2$. The dependence on $\alpha$ of the Majorana phases $\beta_{1,2}$, which are the only sources of low-energy $CP$ violation in the leptonic sector, is shown in the right plot of Fig.~\ref{figure2}.

An important low-energy observable is neutrinoless double beta decay ($0\nu\beta\beta$). In our framework, with no additional sources of flavour violation, its rate is proportional to the modulus of the (11) entry of the effective neutrino mass matrix, denoted by $|m_{ee}|$, in the WB where the charged-lepton mass matrix is diagonal and real. The value of $|m_{ee}|$ is given by
\begin{align}\label{mee1}
|m_{ee}| = \left| m_1\,U_{11}^2+m_2\,U_{12}^2+m_3\,U_{13}^2 \right| \,,
\end{align}
where $U_{ij}$ are the elements of the leptonic mixing matrix $U_\text{PMNS}$. Although with large uncertainties from the poorly known nuclear matrix elements, data available at present set an upper bound on $|m_{ee}|$ in the range 0.2 to 1 eV at 90\% C.L.~\cite{KlapdorKleingrothaus:2000sn,Arnaboldi:2008ds,Wolf:2008hf}. The existing limits will be considerably improved in the forthcoming experiments, with an expected sensitivity of about $10^{-2}$~eV~\cite{Aalseth:2004hb}.

\begin{figure}[h]
\includegraphics[width=7.5cm]{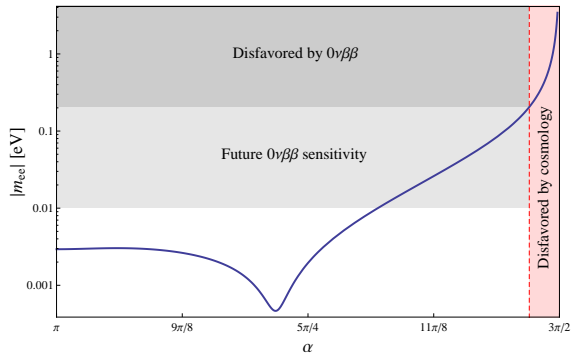}
\caption{\label{figure3} (color online).  Neutrinoless double beta decay parameter $|m_{ee}|$ as a function of $\alpha$.}
\end{figure}

Since in the present model the element $(13)$ is zero in leading order, the only contribution from the Majorana phases to the $0\nu\beta\beta$ decay amplitude will come from the phase $\beta_1$. We may then write Eq.~(\ref{mee1}) as
\begin{align}\label{mee2}
|m_{ee}|= \frac{1}{3}\,\left| 2\,m_1+m_2\,e^{i\beta_1}\right|\,.
\end{align}
In Fig.~\ref{figure3} we can see the evolution of $|m_{ee}|$ as a function of $\alpha$. We obtain $4.66\times 10^{-3}\,\text{eV} \lesssim |m_{ee}| \lesssim 0.20$ eV, where the upper limit comes from imposing the cosmological bound and it corresponds to an almost degenerate neutrino spectrum.

\section{Leptogenesis}
\label{sec4}
Lepton asymmetries produced by out-of-equilibrium decays of heavy neutrinos in the early Universe, at temperatures above $T\sim 10^{12}$ GeV,  do not distinguish lepton flavours. The lepton number asymmetry generated by the $i$-th heavy Majorana neutrino, provided the heavy neutrino masses are far from almost degenerate, would then be given by~\cite{Liu:1993tg}
\begin{align}\label{asym}
\epsilon_i=\frac{1}{8\pi}\,\sum_{j\neq i}\,\frac{\text{Im}
(H_{ij}^2)}{H_{ii}}\,f\left(M^2_j/M^2_i\right)\,,
\end{align}
where
\begin{align}
f(z) = \sqrt{z}
\left[ 1-(1+z)\ln {\frac{1+z}{z}} + \frac{1}{1-z}\right]
\end{align}
and  $H= Y^{\nu\dagger}\,Y^{\nu}$, with $Y^{\nu}$ the Yukawa matrix for the neutrino sector, leading to the Dirac-type neutrino mass matrix, in a WB where $M_R$ is diagonal and real. Notice that $H$ does not depend on whether or not $m_l$ is real and diagonal.

In our framework, if the relations written in Sec.~\ref{sec3} were exact at all energy scales, $H$ would be real and equal to:
\begin{align} \label{Hmatrix}
H&=\frac{M}{v^2}
\begin{pmatrix}
x^2+y^2& 0 &2\,x\,y\,\cos\alpha\\
0 &x^2& 0\\
2\,x\,y\,\cos\alpha& 0 &x^2+y^2
\end{pmatrix}\,.
\end{align}
Therefore, all $\text{Im}(H_{ij}^2)$ would vanish and unflavoured leptogenesis could not take place. Furthermore, the heavy neutrino masses would be exactly degenerate, thus preventing leptogenesis to occur. Flavoured leptogenesis becomes viable once we lift the degeneracy of the heavy Majorana neutrino masses. This is due to the fact that flavoured leptogenesis is sensitive to additional sources of $CP$ violation, as can be seen from the formula for the corresponding $CP$ asymmetry, $\epsilon_i^\alpha$, written below. Notice also that flavoured leptogenesis requires $M \leq 10^{12}$~GeV. From the definition given in Eq.~(\ref{xy}) and Fig.~\ref{figure1} (where we see that $x \lesssim 0.45$) we are able to estimate, if we require this bound on $M$ to be verified, that
\begin{align}
\left|f^u\right| \lesssim 0.08.
\end{align}
This condition for the effective Yukawa couplings is more restrictive than just the need to be in the perturbative regime.

For an almost degenerate heavy Majorana neutrino mass spectrum, leptogenesis can be naturally implemented in the so-called resonant leptogenesis framework~\cite{Pilaftsis:2005rv,Xing:2006ms}. In this case, the $CP$ asymmetry generated by the $i$-th heavy Majorana neutrino decaying into a lepton flavour $\alpha$ is dominated by the one-loop self-energy contributions so that~\cite{Pascoli:2006ci}
\begin{align}\label{resonantCP}
    \epsilon_i^\alpha \simeq -\frac{1}{8\pi}\sum_{j\neq i}\frac{M_i M_j\, \Delta M_{ij}^2}{(\Delta M_{ij}^2)^2+ M_i^2 \Gamma_j^2}\,\frac{\text{Im}[H_{ij}
Y^{\nu *}_{\alpha i} Y^{\nu}_{\alpha j}]}{H_{ii}}\,,
\end{align}
where $\Delta M_{ij}^2=M_j^2-M_i^2$ and $\Gamma_j=H_{jj}\,M_j/(8\pi)$. Defining the mass splitting parameters
\begin{align}\label{deltaNij}
    \delta_{ij}^R = \frac{M_j}{M_i}-1\,,
\end{align}
the $CP$ asymmetries (\ref{resonantCP}) can be conveniently rewritten in the form
\begin{align}\label{resonantCP1}
    \epsilon_i^\alpha \simeq -\frac{1}{16\pi}\sum_{j\neq i}\frac{\delta_{ij}^R}{(\delta_{ij}^R)^2+ \left(\frac{H_{jj}}{16\pi}\right)^2}\,\frac{\text{Im}[H_{ij} Y^{\nu *}_{\alpha i} Y^{\nu}_{\alpha j}]}{H_{ii}}\,.
\end{align}
Notice that when the mass splitting $\delta_{ij}^R$ and the Yukawa matrix $Y^{\nu}$ are independent quantities, $\epsilon_i^\alpha$ is resonantly enhanced for
\begin{align}\label{resonantdelta}
    \delta_{ij}^R \simeq \frac{H_{jj}}{16\pi}\,,
\end{align}
implying~\cite{Pascoli:2006ci}
\begin{align}\label{resonantCP2}
    \epsilon_{i,\text{res}}^\alpha \simeq -\frac{1}{2}\sum_{j\neq i}\frac{\text{Im}[H_{ij} Y^{\nu*}_{\alpha i} Y^{\nu}_{\alpha j}]}{H_{ii} H_{jj}}\,.
\end{align}
In such a case, the $CP$ asymmetry is independent (up to RG running effects) of the absolute heavy Majorana neutrino mass scale $M$.

In Ref.~\cite{Branco:2001pq}, WB invariant $CP$-odd conditions sensitive to the presence of $CP$ violation required for leptogenesis were derived. This type of conditions are a powerful tool for model building since they can be applied to any model without the need to go to a special basis. In the case of unflavoured leptogenesis the $CP$ asymmetry is only sensitive to phases appearing in the matrix $H$ and the relevant WB invariant conditions are given by
\begin{align}\label{wbinv}
I_1 &\equiv {\rm Im Tr}[H M_R^{\dagger}M_R M_R^* H^* M_R]=0, \nonumber\\
I_2 &\equiv {\rm Im Tr}[H (M_R^{\dagger}M_R)^2 M_R^* H^* M_R] = 0,\\
I_3 &\equiv {\rm Im Tr}[H (M_R^{\dagger}M_R)^2 M_R^* H^* M_RM_R^{\dagger}M_R ] = 0.\nonumber
\end{align}
For flavoured leptogenesis, the phases appearing in $H$ are also relevant. There is however still the possibility of generating the required $CP$ asymmetry even for $H$ real. In this case, additional $CP$-odd WB invariant conditions are required since those written above cease to be necessary and sufficient. A simple choice are the WB invariants ${\bar I}_i\, (i=1,2,3)$, obtained from $I_i$ through the substitution of $H$ by ${\bar H}={Y^\nu}^{\dagger} h_\ell Y^\nu$, where $h_\ell=Y^\ell {Y^\ell}^{\dagger}$. For instance, one has~\cite{Branco:2001pq}
\begin{equation}
\bar{I_1}={\rm Im Tr}({Y^\nu}^{\dagger}h_\ell Y^\nu M_R^{\dagger}M_R
M_R^* {Y^\nu}^T h_\ell^* {Y^\nu}^* M_R ),
\label{ibar}
\end{equation}
and similarly for $\bar{I_2}$ and $\bar{I_3}$. As it was the case for $I_i$, $CP$ invariance requires that $\bar{I_i}=0$. The latter $CP$-odd WB invariant conditions are sensitive to the additional phases appearing in flavoured leptogenesis. The well-known Casas-Ibarra parametrization~\cite{Casas:2001sr} makes it clear that the matrix $U_\text{PMNS}$ cancels in $H$. Such is not the case for $Y^{\nu *}_{\alpha i} Y^{\nu}_{\alpha j}\,$. Therefore, flavoured leptogenesis is sensitive to $CP$ violation present at low energies even without any constraints imposed from flavour symmetries. In the case of unflavoured leptogenesis such a connection can only be established in specific flavour models.

In the next subsection we show that the running of parameters from the scale of the $A_4 \times Z_4$ breaking to the scale of the heavy neutrino masses leads to the breaking of the exact degeneracy
of the heavy neutrinos. We then study the case of radiative flavoured leptogenesis, where the mass splitting is generated through renormalization group effects. In the flavoured case, leptogenesis depends on $Y^{\nu}$ computed in the WB where both $M_R$ and $m_l$ are diagonal, since in this case the final charged lepton is well defined, with no summation done. This brings in additional $CP$ violating sources. Notice that $Y^{\nu}$ is proportional to $m_D$ and Eq.~(\ref{mdu}) shows that the matrix $U_\omega$ appears in $Y^{\nu}$ in this WB,
\begin{align}\label{matrixY}
   Y^{\nu}=\frac{\sqrt{M}}{\sqrt{3}\,v}\left(
 \begin{array}{ccc}
 x+y\, e^{i\,\alpha} & x & x+y\, e^{i\,\alpha}\\
 x+\omega y\, e^{i\,\alpha} &\;\omega^2 x\; & y\, e^{i\,\alpha}+\omega x \\
 x+\omega^2 y\, e^{i\,\alpha} & \omega x & y\, e^{i\,\alpha}+\omega^2 x \\
 \end{array}
 \right).
\end{align}

\subsection{Radiative Leptogenesis}

\begin{figure}[t]
\includegraphics[width=7cm]{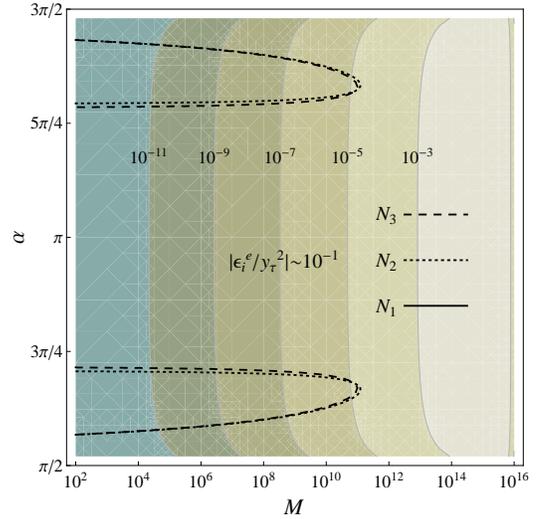}
\caption{\label{figure4} (color online).  The radiatively induced mass splitting (shaded contours) and maximal flavoured $CP$ asymmetries $\epsilon_i^e$ (line contours) in the $(M,\alpha)$ plane.}
\end{figure}

\begin{figure}[t]
\includegraphics[width=7cm]{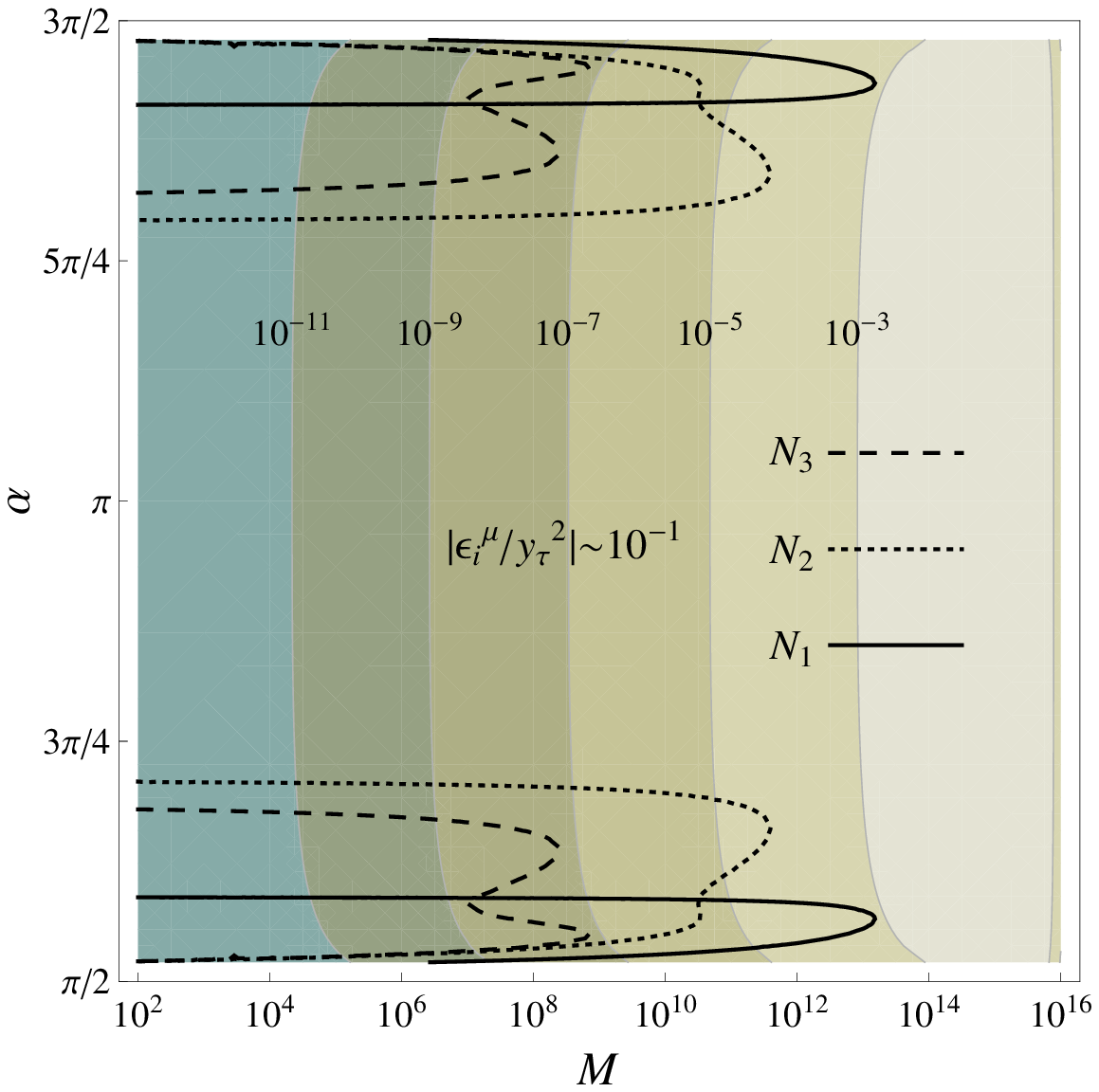}
\caption{\label{figure5} (color online).  The radiatively induced mass splitting (shaded contours) and maximal flavoured $CP$ asymmetries $\epsilon_i^\mu$ (line contours) in the $(M,\alpha)$ plane.}
\end{figure}

\begin{figure}
\includegraphics[width=7cm]{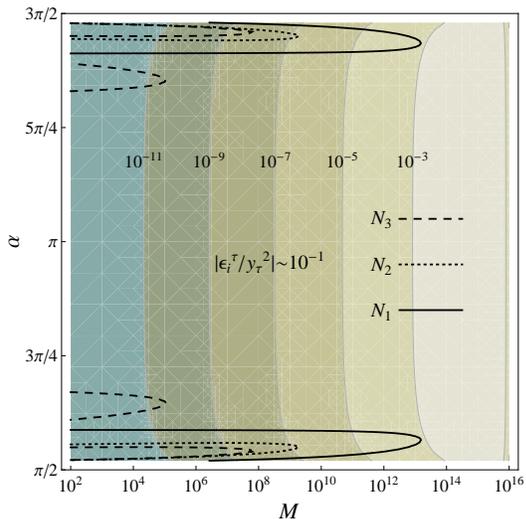}
\caption{\label{figure6} (color online).  The radiatively induced mass splitting (shaded contours) and maximal flavoured $CP$ asymmetries $\epsilon_i^\tau$ (line contours) in the $(M,\alpha)$ plane.}
\end{figure}

Radiative effects due to the renormalization group running from high to low scales can naturally lead not only to a heavy Majorana mass splitting, but also to nonvanishing off diagonal terms in the matrix $H$, which are necessary ingredients for a successful resonant leptogenesis mechanism. In the present framework, the mass splitting generated through the relevant RGE is given by~\cite{GonzalezFelipe:2003fi,Turzynski:2004xy,Branco:2005ye}
\begin{align}\label{deltagen}
\delta_{ij}^R = 2 (\hat{H}_{ii}-\hat{H}_{jj})\, t,\quad
    t = \frac{1}{16\pi^2}\ln \left(\frac{\Lambda^\prime}{M}\right),
\end{align}
where $\hat{H}=V H V^T$ and $V$ is defined in Eq.~(\ref{matrixV}). The cutoff scale $\Lambda^\prime$ is chosen to be equal to the $A_4 \times Z_4$ symmetry breaking scale and close to the GUT scale, $\Lambda^\prime \sim 10^{16}$~GeV. From the form of the matrix $H$ in Eq.~(\ref{Hmatrix}), we then find
\begin{align}\label{deltagen1}
    \delta_{12}^R &= \frac{2M}{v^2}\,\left(y^2-2xy\cos\alpha\right)\, t\,,\nonumber\\
    \delta_{23}^R  &= -\frac{2M}{v^2}\,\left(y^2+2xy\cos\alpha\right)\, t\,,\\
    \delta_{13}^R &= -\frac{8M}{v^2}\,xy\cos\alpha\, t\,.\nonumber
\end{align}
Notice however that a nonvanishing $CP$ asymmetry also requires $\text{Im}[\hat{H}_{ij} \hat{Y}^{\nu*}_{\alpha i} \hat{Y}^{\nu}_{\alpha j}] \neq 0$ with $\hat{Y}^{\nu}=Y^{\nu}\,V^T$ and $Y_\nu$ defined in Eq.~(\ref{matrixY}). Therefore, to have a viable radiative leptogenesis we need to induce nonvanishing $\hat{H}_{ij}\, (i\neq j)$ elements at the leptogenesis scale. This is indeed possible since RG effects due to the $\tau$-Yukawa charged-lepton contribution imply in leading order~\cite{GonzalezFelipe:2003fi,Turzynski:2004xy,Branco:2005ye}
\begin{align}\label{Hijgen}
    \hat{H}_{ij} &\simeq 3 y_\tau^2\,\hat{Y}^{\nu*}_{3 i} \hat{Y}^{\nu}_{3 j}\,t.
\end{align}
The $CP$ flavoured asymmetries can then be obtained from Eqs.~(\ref{resonantCP1}), (\ref{deltagen1}) and (\ref{Hijgen}).

The radiatively induced $CP$ asymmetries $\epsilon_i^\alpha$ are shown in Figs.~\ref{figure4}-\ref{figure6}. Each plot contains two types of contours. The contours represented by lines (solid, dotted and dashed) correspond to the maximum allowed ratio $|\epsilon_i^\alpha/y_\tau^2|\simeq 10^{-1}$ for the decay of each of the three heavy neutrinos into a certain lepton flavour $\alpha$. The color gradient contours are representative of the size of the radiatively induced mass splitting, chosen for illustration to be equal to $\delta_{12}^R$ in all figures. Each contour is depicted as a function of the phase $\alpha$ and the heavy neutrino mass scale $M$. We notice that for temperatures below $10^{12}$~GeV, where flavoured leptogenesis is effective, the induced mass splitting is $\lesssim 10^{-5}$. Such values are sufficiently small to enhanced the $CP$ asymmetries up to values $|\epsilon_i^\alpha| \sim 10^{-5}$ (assuming $y_\tau \sim 10^{-2}$), which in turn can easily lead to the required baryon asymmetry $\eta_B = n_B/n_\gamma \simeq 6.1 \times 10^{-10}$, even for washout factors of the order of $10^{-3}$. We also remark that for temperatures in the range $10^9 \lesssim T \lesssim 10^{12}$~GeV it suffices to consider the leptonic asymmetry $\epsilon_i^\tau$, since in this temperature window only the $\tau$-Yukawa coupling is in thermal equilibrium and $\epsilon_i^e + \epsilon_i^\mu = -\epsilon_i^\tau$. Below $T \sim 10^9$~GeV, all charged-lepton flavours are distinguishable and each asymmetry should be independently considered.

\subsection{Leptogenesis through soft breaking}

In this section we explore the possibility of implementing the mechanism of resonant leptogenesis through a soft breaking of the $A_4$ symmetry at the Lagrangian level. To be specific and simplify our discussion, we shall introduce a single soft-breaking term of the form $\delta M \,\overline{\nu^c_{3R}}\nu^c_{3R}$ \cite{Ma:2001dn} in the Lagrangian of Eq.~(\ref{LagraY}). This term modifies the right-handed neutrino mass matrix and, in turn, its inverse matrix, parametrized here as
\begin{align} \label{MRinv}
M_R^{-1}=\frac{1}{M}
\begin{pmatrix}
1&&\\
&1&\\
&&1+\rho\,e^{i\varphi}
\end{pmatrix}\,,
\end{align}
where the complex number $\rho\,e^{i\varphi}$ characterizes the soft breaking.
The effective neutrino mass matrix obtained through the seesaw mechanism now reads
\begin{align}
m_\nu
= V\,K\,\mathcal{M}_\rho\,K\,V^T,
\end{align}
with
\begin{align}\label{Mmatrix}
\mathcal{M}_\rho =
\begin{pmatrix}
m_1\left(1+\dfrac{\rho}{2}\,e^{i\varphi}\right)&0&\sqrt{m_1m_3}\,\dfrac{\rho}{2}\,e^{i\varphi}\\
0&m_2&0\\
\sqrt{m_1m_3}\,\dfrac{\rho}{2}\,e^{i\varphi}&0&m_3\left(1+\dfrac{\rho}{2}\,e^{i\varphi}\right)
\end{pmatrix},
\end{align}
and the parameters $m_i$ defined in Eq.~(\ref{masses}).

\begin{figure}[h]
\includegraphics[width=7.5cm]{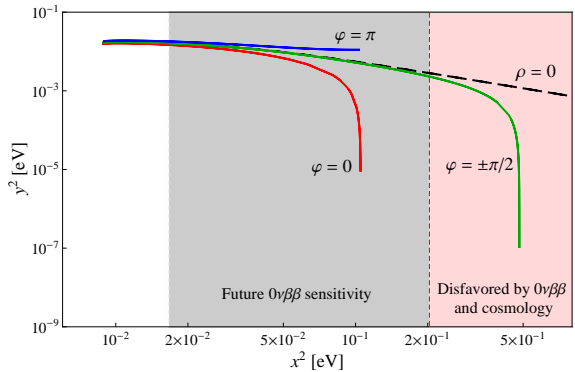}
\caption{\label{figure7} (color online).  The parameter region in the case that the $A_4$ symmetry is softly broken. We take $\rho=0.1$ and consider different values for the soft-breaking phase $\varphi$. The dashed line ($\rho=0$) corresponds to the curve depicted in the right plot of Fig.~\ref{figure1}.}
\end{figure}

The matrix $\mathcal{M}_\rho$ can be diagonalized by the rotation matrix
\begin{align}\label{Vmatrix}
V_\rho=
\begin{pmatrix}
c_\theta&0&s_\theta\,e^{-i\phi}\\
0&1&0\\
-s_\theta\,e^{i\phi}&0&c_\theta
\end{pmatrix},
\end{align}
with $c_\theta \equiv \cos\theta, s_\theta \equiv \sin\theta$,
\begin{align}\label{phi}
\phi=-\text{arctan}\left[\frac{\left(m_3-m_1\right)\sin\varphi}{\left(\rho/2 +\cos\varphi\right)\left(m_3+m_1\right)}\right]\,,
\end{align}
and
\begin{align}\label{theta}
\tan 2\theta = \frac{\rho\sqrt{m_1\,m_3}}{\eta\left(m_3^2-m_1^2\right)}
\left|m_1\,e^{-i\varphi}+m_3\,e^{i\varphi}+\frac{\rho}{2}(m_1+m_3)\right|\,.
\end{align}
In the above expression,
\begin{align}
\eta = 1+\rho\,\cos\varphi+\frac{\rho^2}{4}\,.
\end{align}

\begin{figure}[h]
\includegraphics[width=7.5cm]{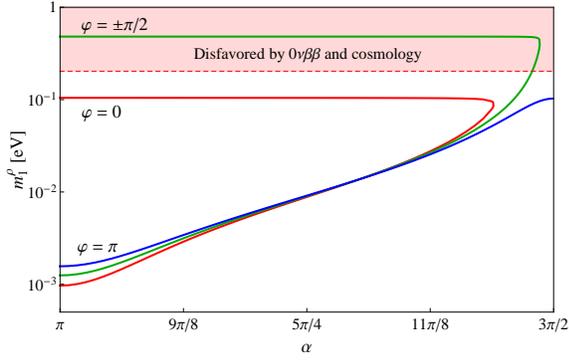}
\caption{\label{figure8} (color online).  The lightest neutrino mass as a function of the high-energy $CP$-violating phase $\alpha$ when $\rho=0.1$ and $\varphi=0,\,\pm \pi/2$ and $\pi$.}
\end{figure}

\begin{figure}
\includegraphics[width=7.5cm]{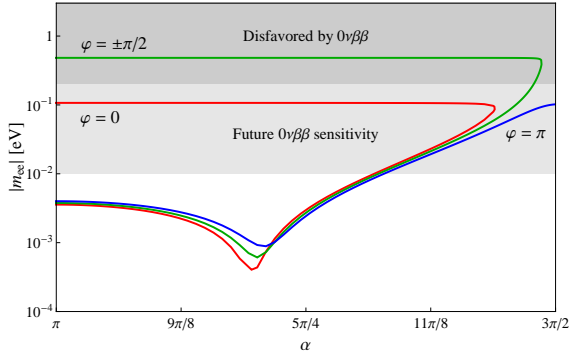}
\caption{\label{figure9} (color online).  Neutrinoless double beta decay parameter $|m_{ee}|$ as a function of $\alpha$ in the case that the symmetry $A_4$ is softly broken. We have taken $\rho=0.1$ and consider $\varphi=0,\,\pm \pi/2$ and $\pi$.}
\end{figure}

The light neutrino masses are given in this case by
\begin{widetext}
\begin{align}\label{masssoft}
\nonumber\left(m_1^\rho\right)^2&= m_1^2 \,\eta+\frac{1}{2}\left[\left(m_3^2-m_1^2\right)\eta+m_1m_3\frac{\rho^2}{2}-\left(\Delta m^2_\text{atm}+\Delta m^2_\text{sol}\right)\right]\,,\\
\left(m_2^\rho\right)^2&= m_2^2\,,\\
\nonumber\left(m_3^\rho\right)^2&= m_3^2 \,\eta -\frac{1}{2}\left[\left(m_3^2-m_1^2\right)\eta-m_1m_3\frac{\rho^2}{2}-\left(\Delta m^2_\text{atm}+\Delta m^2_\text{sol}\right)\right]\,.
\end{align}
\end{widetext}

Notice that there are now five free parameters to be constrained. Besides $x$, $y$ and $\alpha$, already present in $m_i$, two new soft-breaking parameters, $\rho$ and $\varphi$, also appear in Eqs.~(\ref{masssoft}). To further simplify our discussion and to illustrate the main features of the present case, in what follows we assume $\rho=0.1$ and consider $\varphi=0,\,\pm\pi/2\,,\pi$. The allowed parameter region is presented in Fig.~\ref{figure7}. For comparison, the case without soft breaking, i.e. when $\rho=0$, is also plotted (dashed line). The three solid curves correspond to different values of $\varphi$: real positive ($\varphi=0$) and negative ($\varphi=\pi$) soft breaking and a purely imaginary soft breaking ($\varphi=\pm \pi/2$). We note that, for a real value of the soft-breaking parameter, neither the present cosmological bound nor the constraints on $0\nu\beta\beta$ yield a bound more restrictive than the one already imposed by neutrino oscillation data. On the other hand, for $\varphi=\pm \pi/2$, as in the $\rho=0$ case, there is a large region disfavoured by $0\nu\beta\beta$ and cosmology (light area).

Since Eq.~(\ref{masssoft}) does not change the physical meaning of the parameter $x$, namely, $x^2=m_2=m_2^\rho$, the bounds observed in Fig.~\ref{figure7} for $x^2$ and $y^2$
are easily explained, noticing that Eq.~(\ref{atmsol}) now reads
\begin{align}\label{atmsolsoft}
\left(m_1^2+m_3^2\right)\,\eta+m_1m_3\frac{\rho^2}{2}-2m_2^2&=\Delta m^2_\text{atm}-\Delta m^2_\text{sol}\,.
\end{align}
There are two interesting limits arising from this relation. The limit $\alpha=\pi$ was already studied for the case without soft breaking, and can be straightforwardly analyzed in the present case by substituting $\Delta m^2_\text{atm}\rightarrow\Delta m^2_\text{atm}/\eta$ in Eq.~(\ref{y2}). The dependence of $y^2$ on $\rho$ and $\varphi$ would then explain the small splitting between the various curves in Fig.~\ref{figure7}, leading to the relations $y^2(\varphi=0)<y^2(\varphi=\pm\pi/2)<y^2(\varphi=\pi)$.
The second limit, $y\rightarrow0$, is new and leads to completely different phenomenological predictions. In this limit one gets the approximate expression
\begin{align}\label{xsoft}
x^2&\simeq\sqrt{\frac{\Delta m^2_\text{atm}}{2\rho\cos\varphi+\rho^2}}\,.
\end{align}
For $\varphi=0$ we have $x^2\sim 0.1$~eV, while for $\varphi=\pm\pi/2$ the contribution comes only from the second order term in $\rho$ and gives $x^2\sim 0.5$~eV, which is clearly disfavoured by the $0\nu\beta\beta$ decay and cosmological data. Notice also that the above limit is not valid for $\varphi=\pi$, as can be seen from Eq.~(\ref{xsoft}). Nevertheless, the right end point of the $\varphi=\pi$ curve can be estimated from Eq.~(\ref{atmsolsoft}) since it corresponds to $\alpha=\pi/2$ or $3\pi/2$.

The lightest neutrino mass $m_1^\rho$ is plotted in Fig.~\ref{figure8} as a function of the phase $\alpha$. There are two distinct phenomenological regions: one similar to the case without soft breaking and a second one where the light neutrino masses have constant values (with respect to $\alpha$) for a fixed value of $\varphi$ in the range $-\pi/2 \leq \varphi \leq \pi/2$. The latter region is obtained in the limit $y\rightarrow 0$, and corresponds to the vertical branches in Fig.~\ref{figure7} (shown for $\varphi=0$ and $\pm \pi/2$). In this limit, the light neutrino masses are almost degenerate and $m_1^\rho\simeq x^2$. Clearly, for $\varphi=\pm\pi/2$ this region is disfavoured by $0\nu\beta\beta$ and cosmological data. The splitting of the mass for the various values of $\varphi$ when $\alpha=\pi$ is easily understood through the use of Eq.~(\ref{y2}) with the redefinition $\Delta m^2_\text{atm}\rightarrow\Delta m^2_\text{atm}/\eta$.

After diagonalizing the matrix $\mathcal{M}_\rho$ given in Eq.~(\ref{Mmatrix}), the leptonic mixing matrix can be found,
\begin{widetext}
\begin{align}\label{PMNSsoft}
&\nonumber U_\text{PMNS}=U^\dagger\left(\omega\right)V K^\prime V_\rho\\
&= e^{i\pi/2}\begin{pmatrix}
1&&\\
&\omega^2&\\
&&\omega
\end{pmatrix}\left[
\begin{pmatrix}
\sqrt{\frac{2}{3}}\,e^{i\sigma_1}\,c_\theta&\frac{1}{\sqrt{3}}&0\\
-\frac{1}{\sqrt{6}}\,e^{i\sigma_1}\,c_\theta&\frac{1}{\sqrt{3}}&-\frac{1}{\sqrt{2}}\,e^{i\sigma_2}\,c_\theta\\
-\frac{1}{\sqrt{6}}\,e^{i\sigma_1}\,c_\theta&\frac{1}{\sqrt{3}}& \frac{1}{\sqrt{2}}\,e^{i\sigma_2}\,c_\theta
\end{pmatrix} +s_\theta
\begin{pmatrix}
0&0&\sqrt{\frac{2}{3}}\,e^{i\phi}\\
\frac{1}{\sqrt{2}}\,e^{i(\sigma_2+\phi)}&0&-\frac{1}{\sqrt{6}}\,e^{i(\sigma_1-\phi)}\\
-\frac{1}{\sqrt{2}}\,e^{i(\sigma_2+\phi)}&0& -\frac{1}{\sqrt{6}}\,e^{i(\sigma_1-\phi)}
\end{pmatrix} \right],
\end{align}
\end{widetext}
and the remaining low-energy observables determined. The effective mass parameter relevant for $0\nu\beta\beta$ decay [cf. Eq.~(\ref{mee1})] is presented in  Fig.~\ref{figure9} for different values of $\varphi$. The analysis of the plot is similar to the one of Fig.~\ref{figure8}. Once again, there are two distinct regions. In particular, when $y\rightarrow0$ and  $-\pi/2 \leq \varphi \leq \pi/2$, the effective mass parameter $|m_{ee}|$ tends to a constant value given by $|m_{ee}|\simeq x^2$.

\begin{figure}[t]
\includegraphics[width=7.5cm]{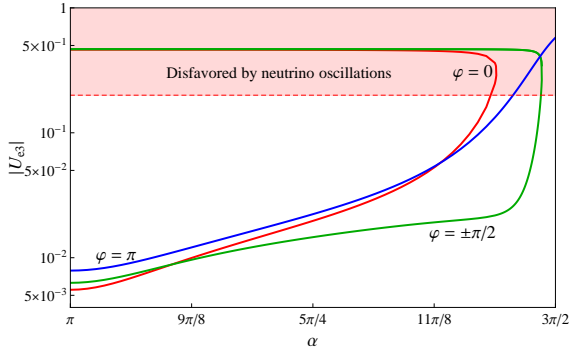}
\caption{\label{figure10} (color online).  The absolute value of the element $U_{e3}$ of the PMNS mixing matrix as a function of $\alpha$ when $\rho=0.1$ and $\varphi=0,\,\pm \pi/2$ and $\pi$.}
\end{figure}

\begin{figure}[t]
\includegraphics[width=7.5cm]{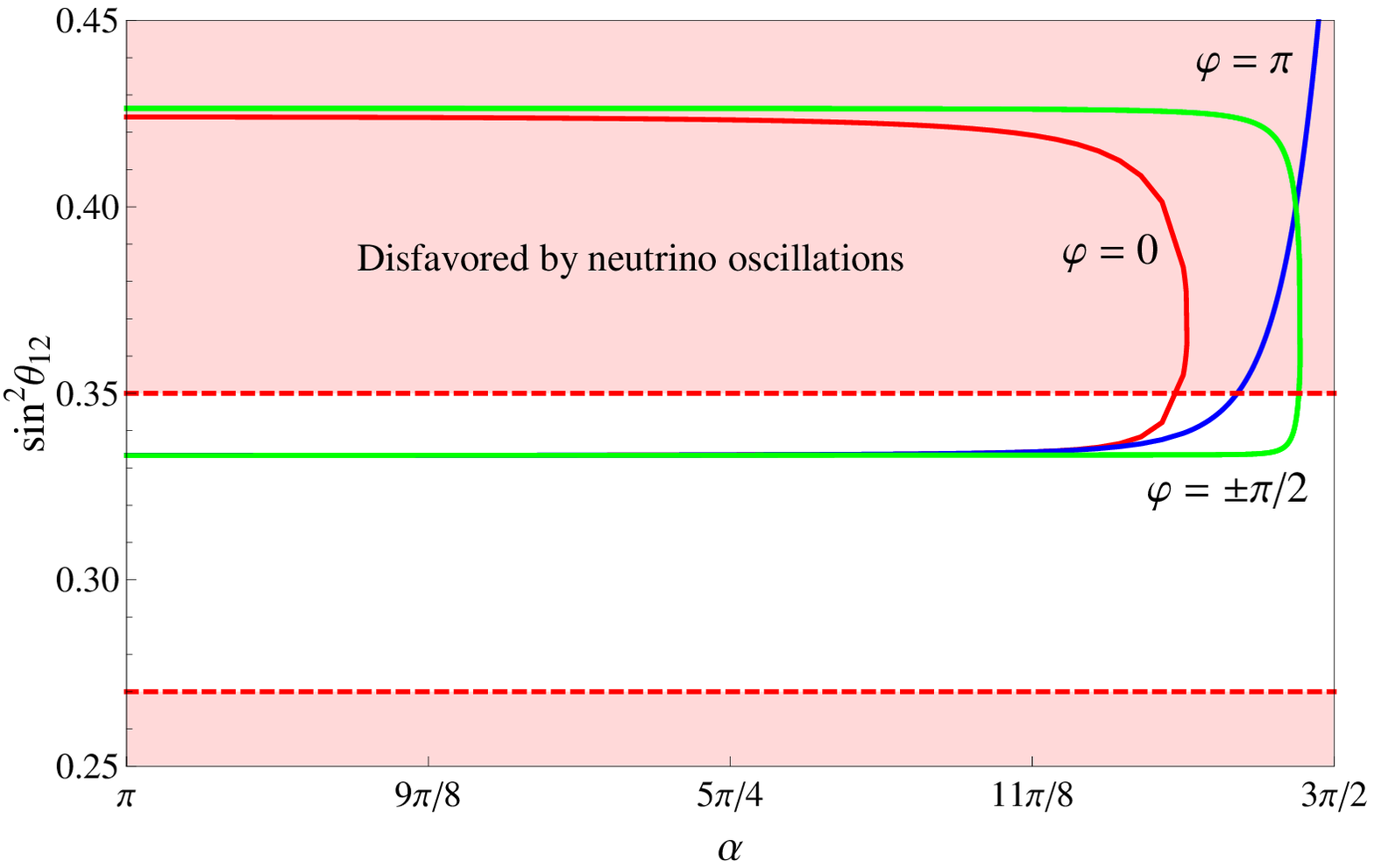}
\caption{\label{figure10a} (color online).  $\sin^2 \theta_{12}$ as a function of $\alpha$ for $\rho=0.1$ and $\varphi=0,\,\pm \pi/2$ and $\pi$.}
\end{figure}

\begin{figure}[t]
\includegraphics[width=7.5cm]{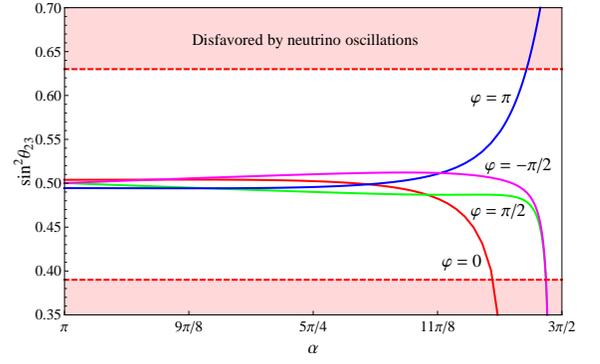}
\caption{\label{figure10b} (color online).  $\sin^2 \theta_{23}$ as a function of $\alpha$ for $\rho=0.1$ and $\varphi=0,\,\pm \pi/2$ and $\pi$.}
\end{figure}

Another feature of this case is the prediction of a nonvanishing $U_{e3}$ matrix element. Its absolute value, $\left|U_{e3}\right|$, is plotted in Fig.~\ref{figure10} as a function of $\alpha$ for the various values of $\varphi$. We notice that the phenomenological region that predicts constant values of $\left|U_{e3}\right|$ is already disfavoured by the neutrino oscillation data, which implies the constraint $\left|U_{e3}\right| \lesssim 0.2$ at $2\sigma$ level~\cite{Schwetz:2008er}. Indeed, from Eq.~(\ref{theta}) and in the limit when $y \rightarrow 0$ we get $\theta \simeq \pi/4$, which then yields $\left|U_{e3}\right| \simeq 1/\sqrt{3}$. This upper bound is reduced when the small corrections due to $y$ are taken into account. From Fig.~\ref{figure10} we estimate the maximum value to be $\left|U_{e3}\right|\simeq 0.47$. On the other hand, in the region where $\alpha=\pi$ and $y^2 \simeq x^2$ we get
\begin{align}
\left|U_{e3}\right|&\simeq\frac{\rho}{\sqrt{6}}\frac{\sqrt[4]{\Delta m^2_\text{atm}/\eta}-2\sqrt[4]{\Delta m^2_\text{sol}}}{\sqrt[4]{\Delta m^2_\text{atm}/\eta}}\,,
\end{align}
which explains the splitting of the three curves and also leads to the allowed range of values $\left|U_{e3}\right|\sim(5-7)\times 10^{-3}$. The corresponding mixing angles $\theta_{12}$ and $\theta_{23}$ are presented in Figs.~\ref{figure10a} and \ref{figure10b}, respectively. In these figures, the light (red) shaded regions are presently excluded at $2\sigma$ by the global analysis of neutrino oscillation data~\cite{Schwetz:2008er}.

\begin{figure}[t]
\includegraphics[width=7.5cm]{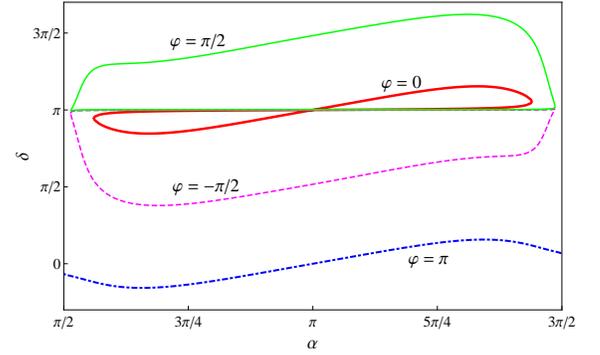}
\caption{\label{figure11} (color online).  The dependence of the low-energy $CP$-violating Dirac phase $\delta$ on the high-energy phase $\alpha$ for different values of the soft-breaking phase $\varphi$ and $\rho=0.1$.}
\end{figure}

Finally, in order to identify the low-energy Dirac phase $\delta$ and the Majorana phases $\beta_{1,2}$, we rewrite the PMNS mixing matrix (\ref{PMNSsoft}) in the standard parametrization~\cite{Amsler:2008zzb}. The following relations hold:
\begin{align}\label{betadirac}
\beta_1=-\sigma_1,\quad \beta_2-\beta_1&=\phi+\delta\,,
\end{align}
with $\phi$ defined in Eq.~(\ref{phi}). We recall that in the limit where there is no soft-breaking term one has $\beta_2-\beta_1=\sigma_2$, which is not obvious from Eq.~(\ref{betadirac}), since the phases $\phi$ and $\delta$ have no physical meaning in this limit. The dependence of the low-energy Dirac phase $\delta$ on the high-energy phase $\alpha$ is shown in Fig.~\ref{figure11} for different values of $\varphi$. We note that $\delta$ is quite sensitive to $\sin \varphi$. The constant lines for $\delta=\pi$ correspond to the vertical branches in Fig.~\ref{figure7}, so that for $\varphi=\pm\pi/2$ they are excluded by the cosmological and $0\nu\beta\beta$ bounds. The dependence of the Majorana phases $\beta_{1,2}$ on the phase $\alpha$ is not much affected by the soft-breaking term and is quite similar to the one shown in Fig.~\ref{figure2}.

\begin{figure*}[t]
\begin{tabular}{cc}
\includegraphics[width=7cm]{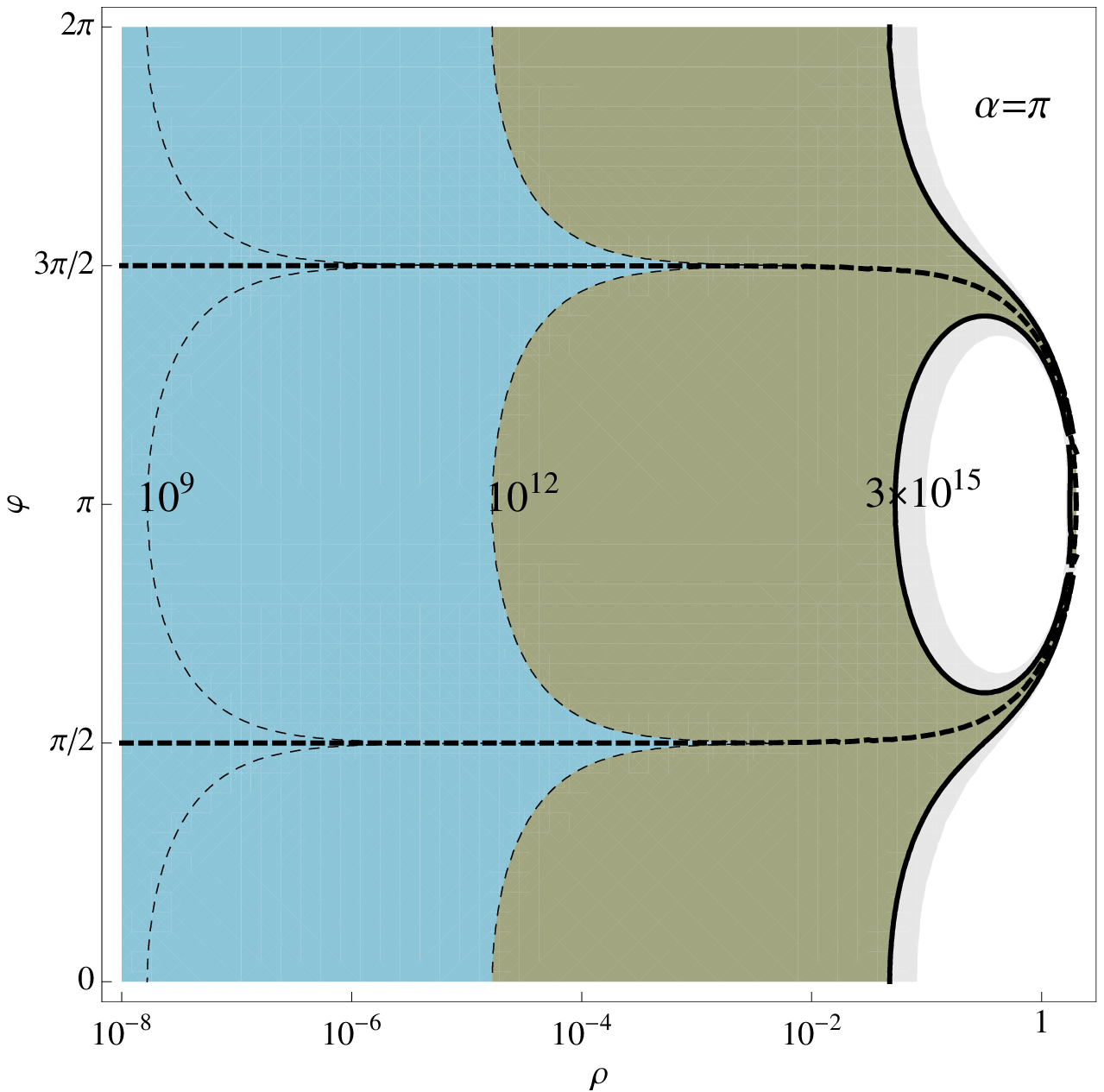}&
\includegraphics[width=7cm]{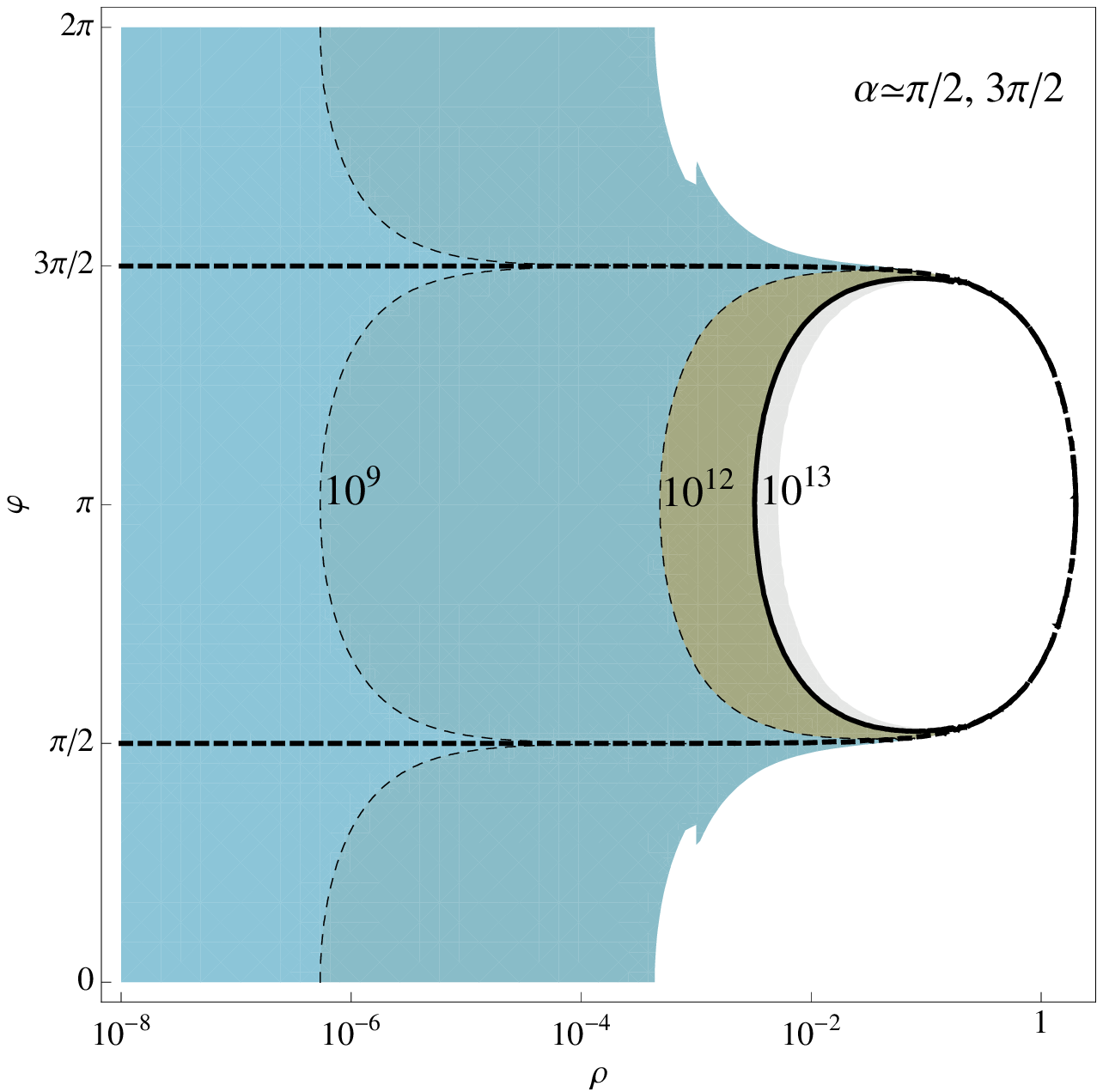}
\end{tabular}
\caption{\label{figure12} (color online).  Contours of constant $M$ (dashed lines) in the $(\rho,\varphi)$ plane for $\alpha=\pi$ (left plot) and $\alpha \simeq \pi/2$ or $3\pi/2$ (right plot). The thick dashed curve corresponds to the fine-tuned region where $\cos\varphi \simeq -\rho/2$.}
\end{figure*}

Let us now analyze the viability of leptogenesis and its possible connection with low-energy neutrino observables. We start by evaluating the Dirac neutrino Yukawa coupling matrix $Y^\nu$ in the basis where the charged leptons and heavy Majorana neutrinos are real and diagonal. In this case, $Y^\nu$ defined in Eq.~(\ref{matrixY}) becomes $Y^\nu\,\text{diag}(1,1,e^{-i\,\gamma/2})$, where $\gamma=-\text{arg}\left(1+\rho\,e^{i\varphi}\right)$ is the phase of the matrix element $\left(M_R\right)_{33}$. The matrix $H= Y^{\nu\dagger}\,Y^{\nu}$ now becomes complex:
\begin{align}\label{Hmatrixsoft}
H&=\frac{M}{v^2}\,
\begin{pmatrix}
x^2+y^2&0&2\,x\,y\,\cos\alpha\,e^{-i\,\gamma/2}\\
0&x^2&0\\
2\,x\,y\,\cos\alpha\,e^{i\gamma/2}&0&x^2+y^2
\end{pmatrix}\,.
\end{align}
Therefore, a crucial difference from the radiative leptogenesis case studied in the previous section is the possibility of having unflavoured leptogenesis. To illustrate its main features, in what follows we restrict our discussion to the resonantly enhanced $CP$ asymmetries given in Eq.~(\ref{resonantCP2}), provided that the condition (\ref{resonantdelta}) is satisfied. This will also allow us to estimate the maximal value of the leptonic asymmetries that can be reached in the present framework.

For small values of $\rho$, the resonant condition given by Eq.~(\ref{resonantdelta}), together with the definition of the mass splitting $\delta_{ij}^R\,(i,j=1,3)$ in Eq.~(\ref{deltaNij}) and the matrix $M_R^{-1}$ in Eq.~(\ref{MRinv}), imply the relation
\begin{align}\label{deltasoft}
|2\rho\,\cos\varphi+\rho^2| \simeq \frac{1}{16\,\pi}\frac{M}{v^2}\left(m_3+m_1\right)\,.
\end{align}

From the above equation we can estimate the heavy neutrino mass scale necessary to resonantly enhance the leptonic asymmetries. We obtain
\begin{align}
M\simeq\left(1.5\times10^{15}\,\text{GeV}\right)\left(\frac{1 \,\text{eV}}{m_3+m_1}\right)|2\rho\,\cos\varphi+\rho^2|\,.
\end{align}
In Fig.~\ref{figure12} we present the contours of constant $M$ in the $(\rho,\varphi)$ plane for $\alpha=\pi$ and close to $\pi/2$ (or to $3\pi/2$). The contour line $M=10^{12}$~GeV sets the transition from unflavoured to flavoured leptogenesis, while $M=10^{9}$~GeV corresponds to the temperature below which the three charged-lepton flavours are distinguishable. As can be seen from the figure, when $\alpha=\pi$ there is a large parameter region where unflavoured resonant leptogenesis could be viable.
On the other hand, a resonantly enhanced flavoured leptogenesis would in general require the soft-breaking parameter $\rho$ to be very small or the fine-tuned relation $\cos\varphi \simeq -\rho/2$ to be satisfied. As $\alpha$ tends to $\pi/2$ (or $3\pi/2$) the unflavoured leptogenesis region shrinks, while the flavoured leptogenesis one shifts to higher values of $\rho$. There is in each case an upper bound on the heavy Majorana neutrino mass: $M \simeq 3\times 10^{15}$~GeV for $\alpha=\pi$ and $M \simeq 10^{13}$~GeV  for $\alpha=\pi/2$ (or $3\pi/2$).

Denoting $\epsilon_{\text{res}}\equiv \epsilon_{1,\text{res}}=-\epsilon_{3,\text{res}}$  ($\epsilon_{2,\text{res}}=0$), the unflavoured $CP$ asymmetry is given in this case by
\begin{align} \label{unflavorcpasym}
\epsilon_{\text{res}}\simeq -\frac{1}{2}\left(\frac{m_3-m_1}{m_3+m_1}\right)^2\rho\sin\varphi,
\end{align}
and attains its maximal value $(\epsilon_{\text{res}})_\text{max} \simeq -0.45\, \rho \sin \varphi$ when neutrinos are hierarchical. For the $e$-flavoured $CP$ asymmetry we find:
\begin{align}
\epsilon^e_{\text{res}}\simeq\frac{1}{3}\frac{\left(m_3-m_1\right)m_1}{\left(m_3+m_1\right)^2}
\,\rho\,\sin\,\varphi\,,
\end{align}
which has the maximal value $(\epsilon^e_{\text{res}})_\text{max} \simeq 4\times 10^{-2} \rho \sin \varphi$. The $\mu$- and $\tau$-flavoured asymmetries are given by
\begin{equation}
\begin{split}
\epsilon^{\mu,\tau}_{\text{res}}\simeq & \frac{1}{12}\frac{m_3-m_1}{\left(m_3+m_1\right)^2}\left[\pm\sqrt{3}\left(4m_2-m_1-m_3\right)\right.\\
-&\left.\left(3m_3-m_1\right)\rho\sin\,\varphi\right]\,,
\end{split}
\end{equation}
which, clearly, are not suppressed by the soft-breaking parameter $\rho$ and can reach values up to $7 \times10^{-2}$ for hierarchical neutrinos.

\begin{figure}[h]
\includegraphics[width=7.5cm]{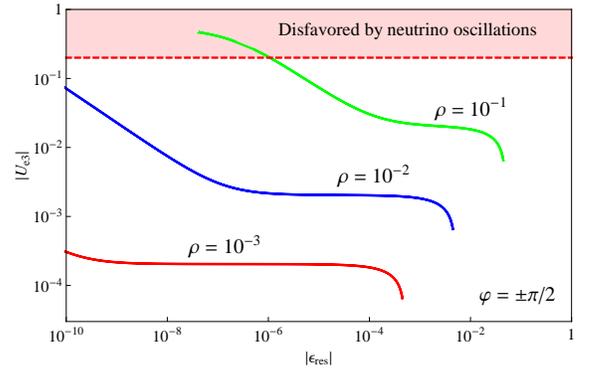}
\caption{\label{figure13} (color online).  The correlation between the mixing matrix element $|U_{e3}|$ and the unflavoured leptonic asymmetry $\epsilon_{\text{res}}$ for $\varphi=\pm \pi/2$ and different values of the soft-breaking values $\rho$.}
\end{figure}

Finally, in Fig.~\ref{figure13} we present the correlation between the low-energy observable $|U_{e3}|$ and the absolute value of the unflavoured leptonic asymmetry $|\epsilon_{\text{res}}|$ for different values $\rho$.  The curves are shown for $\varphi=\pm \pi/2$, which yields the maximal asymmetry [cf. Eq.~(\ref{unflavorcpasym})]. As can be seen from the figure, in the region of phenomenological interest ($|\epsilon_\text{res}| \gtrsim 10^{-6}$) a variation of the soft-breaking parameter $\rho$ simply implies a rescaling of the curves, once both quantities, $|U_{e3}|$ and $|\epsilon_{\text{res}}|$, are proportional to $\rho$ in this region.

\section{Conclusions}
\label{sec5}
Recently, models based on discrete flavour symmetries~\cite{Ma:2001dn,Altarelli:2005yx,He:2006dk,deMedeirosVarzielas:2005qg,Altarelli:2005yp,Luhn:2007sy,Jenkins:2008rb,Yin:2009ic} have attracted much attention due to the possibility of finding implementations that lead to the HPS mixing matrix in leading order. The implications of these symmetries for leptogenesis depend on the specific details of the model. Among these models, those based on type-I seesaw realizations have in general the common prediction of vanishing leptonic $CP$ asymmetries, since the combination $Y^{\nu\dagger} Y^\nu$, relevant for leptogenesis, is proportional to the unit matrix. Thus, higher dimensional operators, suppressed by additional powers of the cutoff scale $\Lambda$ are usually required to allow for leptogenesis in these models~\cite{Jenkins:2008rb,Yin:2009ic}. We have presented an explicit example, based on the $A_4$ symmetry, where the above limitations can be overcome. The model is based on an effective theory with an $A_4 \times Z_3 \times Z_4$ symmetry, which is spontaneously broken at a high scale, leading to exact tribimaximal leptonic mixing in leading order. A particular feature of the model is the degeneracy of the heavy Majorana neutrino mass spectrum. Therefore, for leptogenesis to become viable this degeneracy must be lifted. This can be easily achieved either by renormalization group effects or by a soft breaking of the $A_4$ symmetry, which then naturally leads to a viable resonant leptogenesis mechanism.

We have also studied the implications for low-energy neutrino physics. The model can accommodate a hierarchical or an almost degenerate light neutrino spectrum. It also gives definite predictions for the $0\nu\beta\beta$ decay mass parameter $|m_{ee}|$. In the so-called radiative leptogenesis framework, the HPS mixing pattern is exact up to negligible running effects. Furthermore, only a flavoured leptogenesis regime is allowed. If the $A_4$ symmetry is softly broken, e.g. by a mass term that lifts the heavy Majorana neutrino degeneracy, then both unflavoured and flavoured leptogenesis can be implemented. In this case, corrections to tribimaximal mixing would lead to a nonvanishing $U_{e3}$ and definite predictions for the low-energy $CP$-violating phases.

\end{document}